%% file: iclr2026_conference.tex
\documentclass{article} 
\usepackage{iclr2026_conference,times}

\input{math_commands.tex}

\usepackage{hyperref}
\usepackage{url}

\usepackage{subcaption}
\usepackage{marvosym}
\usepackage{arydshln}
\usepackage{enumitem}
\usepackage{amsmath}
\usepackage{amssymb}
\usepackage{mathtools}
\usepackage{amsthm}
\usepackage{amssymb}
\usepackage{color}
\usepackage{colortbl}
\usepackage{multirow}
\usepackage{makecell}

\usepackage{bbding}
\usepackage{wrapfig}
\usepackage{amsfonts}
\usepackage{pifont} 

\usepackage{diagbox} 
\usepackage{xcolor}
\usepackage{booktabs}
\usepackage{wasysym}
\usepackage{xspace}
\usepackage{hhline}
\newcommand{\ourbench}{{\textsc{STAR-Bench}}\xspace}

\usepackage[capitalise]{cleveref}
\Crefname{section}{Sec.}{Secs.}
\Crefname{table}{Tab.}{Tabs.}
\crefname{appendix}{Appendix}{Appendices}
\Crefname{appendix}{Appendix}{Appendices}

\definecolor{darkblue}{RGB}{94,110,186}
\definecolor{SciGreen}{RGB}{44, 160, 44}      
\definecolor{SciRed}{RGB}{214, 39, 40}       
\definecolor{SciOrange}{RGB}{230, 159, 0}    
\definecolor{myred}{RGB}{220,20,60}

\newenvironment{hlblock}
  {} 
  {}           

\newcommand{\cmark}{\textcolor{SciGreen}{\ding{51}}}
\newcommand{\xmark}{\textcolor{SciRed}{\ding{55}}}
\newcommand{\pmark}{\textcolor{SciOrange}{\LEFTcircle}}
\newcommand{\darkblue}[1]{\textcolor{darkblue}{#1}}

\title{STAR-Bench: Probing Deep Spatio-Temporal Reasoning as Audio 4D Intelligence} 

\author{%
  Zihan Liu$^{1,2}$, Xiaoran Liu$^{1,4}$, Yuhang Zang$^{2}$\textsuperscript{\Letter}, Xiaoyi Dong$^{2}$, \\ 
  \textbf{Yuhang Cao$^{2}$}, \textbf{Jiaqi Wang$^{2,4}$}\textsuperscript{\Letter}, \textbf{Xipeng Qiu$^{1,4}$},
    \textbf{Dahua Lin$^{2,3}$}.  \\
  $^{1}$ Fudan University, $^{2}$ Shanghai AI Laboratory, \\ $^{3}$The Chinese University of Hong Kong, $^{4}$ Shanghai Innovation Institute \\
  {\tt\small xlwei24@m.fudan.edu.cn, zangyuhang@pjlab.org.cn} \\
  {{\tt\small Github: \url{https://github.com/InternLM/SIM-CoT}}}
}
\author{%
  Zihan Liu$^{1,2*}$, Zhikang Niu$^{3,5*}$, Qiuyang Xiao$^{3}$, Zhisheng Zheng$^{3}$,  Ruoqi Yuan$^{1}$, Yuhang Zang$^{2}$\textsuperscript{\Letter}, \\
   \textbf{Yuhang Cao$^{2}$, Xiaoyi Dong$^{2,4}$, Jianze Liang$^{2}$, Xie Chen$^{3,5}$, Leilei Sun$^{1}$, Dahua Lin$^{2,4}$,  
  Jiaqi Wang$^{2,5}$\textsuperscript{\Letter}}  \\
  $^{1}$ Beihang University, $^{2}$ Shanghai AI Laboratory,  $^{3}$ Shanghai Jiao Tong University,  \\
  $^{4}$The Chinese University of Hong Kong, $^{5}$ Shanghai Innovation Institute \\
  {\tt\small liuzihan@buaa.edu.cn, zangyuhang@pjlab.org.cn} \\
  {{\tt\small \textbf{Code}: \url{https://github.com/InternLM/StarBench}}}\\
  {{\tt\small \textbf{Benchmark}: \url{https://huggingface.co/datasets/internlm/STAR-Bench}}}\\
  {{\tt\small \textbf{Homepage}: \url{https://internlm.github.io/StarBench}}}
}

\iclrfinalcopy 
\begin{document}

\maketitle

\input{0-Abstract}
\input{1-Introduction}

\input{2-Related_Work}
\input{3-Method}
\input{4-Experiments}

\input{5-Conclusion}

\newpage

\bibliography{iclr2026_conference}
\bibliographystyle{iclr2026_conference}

\appendix
\input{6-Appendix}

\end{document}

%% file: math_commands.tex
\usepackage{amsmath,amsfonts,bm}

\def\eqref#1{equation~\ref{#1}}

\def\1{\bm{1}}

\DeclareMathAlphabet{\mathsfit}{\encodingdefault}{\sfdefault}{m}{sl}
\SetMathAlphabet{\mathsfit}{bold}{\encodingdefault}{\sfdefault}{bx}{n}



%% file: 0-Abstract.tex
\begin{abstract}
Despite rapid progress in Multi-modal Large Language Models and Large Audio-Language Models, existing audio benchmarks largely test semantics that can be recovered from text captions, masking deficits in fine-grained perceptual reasoning.
We formalize audio \textbf{4D intelligence} that is defined as reasoning over sound dynamics in time and 3D space, and introduce \textbf{STAR-Bench} to measure it.
STAR-Bench combines a Foundational Acoustic Perception setting (six attributes under absolute and relative regimes) with a Holistic Spatio-Temporal Reasoning setting that includes segment reordering for continuous and discrete processes and spatial tasks spanning static localization, multi-source relations, and dynamic trajectories.
Our data curation pipeline uses two methods to ensure high-quality samples.
For foundational tasks, we use procedurally synthesized and physics-simulated audio.
For holistic data, we follow a four-stage process that includes human annotation and final selection based on human performance.
Unlike prior benchmarks where caption-only answering reduces accuracy slightly, STAR-Bench induces far larger drops (-31.5\% temporal, -35.2\% spatial), evidencing its focus on linguistically hard-to-describe cues.
Evaluating 19 models reveals substantial gaps compared with humans and a capability hierarchy: closed-source models are bottlenecked by fine-grained perception, while open-source models lag across perception, knowledge, and reasoning.
Our STAR-Bench provides critical insights and a clear path forward for developing future models with a more robust understanding of the physical world.
\end{abstract}

%% file: 1-Introduction.tex
\section{Introduction}\label{sec:intro}
As a fundamental modality of human perception, audio serves a pivotal role in communication, aesthetic appreciation, and situational awareness, complementing the limitations of visual perception.
With the rise of Multimodal Large Language Models (MLLMs) \citep{gemini25pro,gpt4o} and especially Large Audio-Language Models (LALMs) \citep{Qwen2-Audio,Audio-Flamingo3}, these models have shown impressive capabilities in understanding audio, representing a crucial step toward diverse applications such as embodied intelligence \citep{paul2022avlen}.

\begin{figure*}[tb!]
	\centering
	\includegraphics[width=0.93\columnwidth]{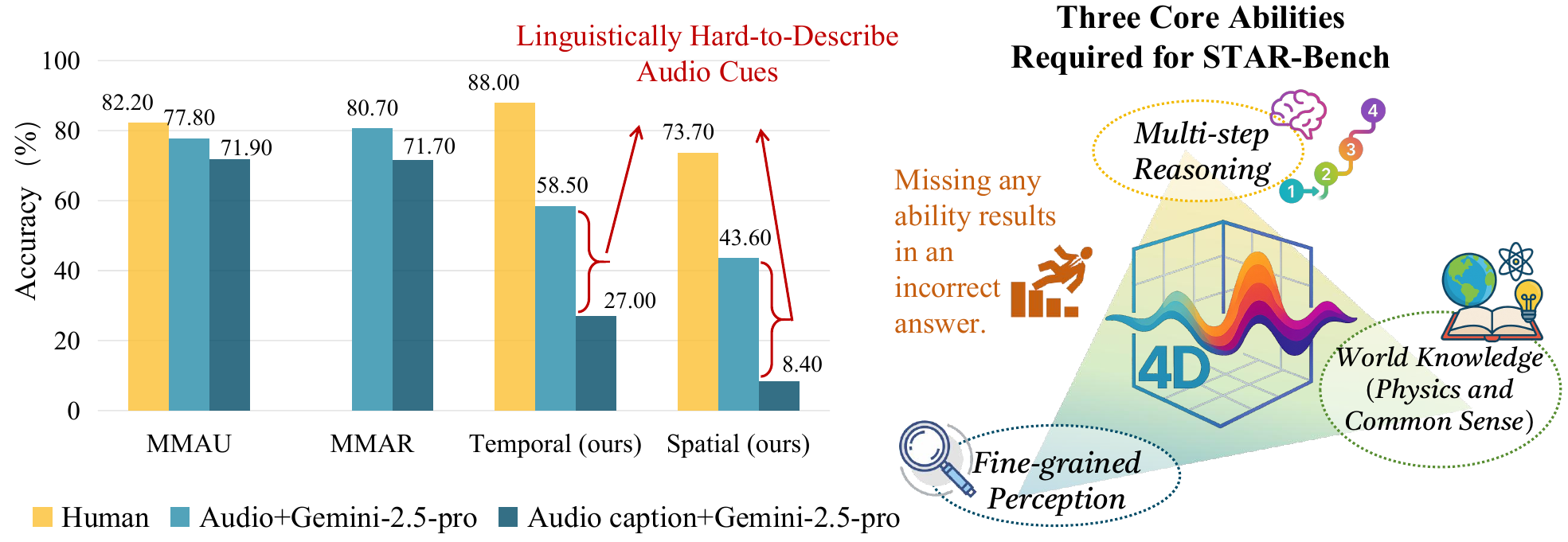}
    \vspace{-6pt}
	\caption{\textbf{(Left)}: A comparison between humans and the Gemini 2.5 Pro with and without audio captions on various audio benchmarks. Our STAR-Bench evaluates linguistically hard-to-describe audio cues. See \cref{appendix:caption_prompt} for audio caption details. \textbf{(Right)}: The three core abilities required to solve tasks in the STAR-Bench benchmark.}
	\label{fig:teaser}
    \vspace{-20pt}
\end{figure*}

To drive progress, a series of audio benchmarks has been introduced \citep{airbench, MMAU}, covering traditional tasks like Automatic Speech Recognition (ASR) and sound event classification.
While some recent efforts are beginning to emphasize reasoning abilities \citep{ma2025mmar, mmaupro}, we observe that existing benchmarks predominantly focus on coarse-grained semantic content, which is audio information that can be distilled into textual descriptions with minimal loss.
As shown in the \textbf{left} part of \cref{fig:teaser}, we first use Gemini 2.5 Pro \citep{gemini25pro} to generate detailed audio captions for samples in recent representative audio benchmarks MMAU (test-mini) \citep{MMAU} and MMAR \citep{ma2025mmar}.
We then prompt the model to answer questions based \textit{only} on these audio captions, and its performance drops by only 5.9\% and 9.0\%, respectively, compared to when it processes the raw audio.
This result suggests that existing benchmarks primarily evaluate audio information that is \textbf{easily representable by text}.
However, human auditory intelligence is not limited to this coarse-grained understanding.
For example, humans can intuitively judge the water level in a container from the dynamic changes in the pouring sound, even without being able to precisely articulate the underlying acoustic features.
Similarly, we can infer the trajectory and distance of a vehicle approaching from behind to ensure our safety.
These abilities are rooted in deep reasoning of audio cues \textbf{that are difficult to represent linguistically}.

To capture this human-like audio competence, we propose a new paradigm, called \textbf{audio 4D intelligence}.
This is defined as the ability to perform deep reasoning over the dynamics of \textbf{sound sources} in \textbf{time (1D)} and \textbf{three-dimensional space (3D)}, grounded in an understanding of the physical world.
Mastering 4D audio intelligence is crucial for various applications.
In embodied AI and robotics, for instance, agents must integrate fine-grained auditory cues to interact naturally with their surroundings, such as using sound to infer the trajectory of an object or to monitor the subtle operations of a machine.
To systematically evaluate this paradigm and bridge the gap between current audio benchmarks and real-world auditory intelligence, we introduce the \textbf{S}patio-\textbf{T}emporal \textbf{A}udio \textbf{R}easoning (\textbf{\ourbench}) benchmark.

\ourbench is designed through a hierarchical task structure with two levels.
At the \textbf{Foundational Acoustic Perception} level, we conduct a fine-grained, quantitative evaluation of six core audio attributes (pitch, loudness, duration, azimuth, elevation, distance) across both absolute perception ranges and relative discrimination sensitivity.
We also introduce a \textbf{Holistic Spatio-Temporal Reasoning} level that evaluates an audio model’s ability to infer both event order and 3D scene structure.
Temporal reasoning is tested via segment reordering that spans continuous processes and discrete event scripts, while spatial reasoning covers static localization, multi-source relations, and dynamic trajectory tracking.
As shown in the \textbf{right} part of \cref{fig:teaser}, every question in our holistic tasks is designed to probe a synthesis of three core pillars, such as multi-step reasoning.
A failure in any one of these pillars will lead to an incorrect response.
Our \textbf{data curation pipeline} couples procedurally synthesized, fully parameterized audio for foundational perception with large-scale real-world corpora for holistic reasoning.
For the latter, we use a four-stage process including \textbf{human annotation} and \textbf{final selection by human performance} to ensure the high quality of benchmark samples.

Our comprehensive evaluation of 19 models (16 open-source and 3 closed-source) reveals a clear capability hierarchy between the two groups.
Leading closed-source models like Gemini 2.5 Pro excel in knowledge and reasoning, shifting their primary bottleneck to the more difficult challenge of fine-grained perception. In contrast, open-source models exhibit fundamental weaknesses across all three core capabilities. Through our detailed error analysis and ablation studies, we highlight several key insights for the future development of open-source audio models:
1) \textbf{Enhancing dense audio captioning.} Open-source models struggle to produce dense, fine-grained captions, which limits their perceptual sensitivity and ability to extract embedded knowledge. Bridging this gap is a crucial first step.
2) \textbf{Improving multi-audio reasoning.} Open-source models lag significantly in comparing, integrating, and grounding information across multiple audio clips.
3) \textbf{Moving beyond channel-averaged audio preprocessing.}  The common practice of averaging multi-channel audio into a mono signal is a major bottleneck for spatial reasoning. Developing architectures that natively process multi-channel cues is essential for unlocking genuine spatial awareness.

Our contributions are summarized as: \textbf{(1)} We formalize \textbf{audio 4D intelligence}, and empirically show that prior benchmarks largely probe text-representable semantics, motivating a shift toward fine-grained, non-linguistic auditory cues.
\textbf{(2)} We introduce the \ourbench with foundational acoustic perception and holistic spatio-temporal reasoning tasks, together with a rigorous curation pipeline with expert validation.
\textbf{(3)} We provide a comprehensive evaluation of 19 LALMs/OLMs. Our analyses and standardized protocols establish strong baselines and testbeds for future research.

%% file: 2-Related_Work.tex
\section{Related Work}
The recent progress of Large Audio-Language Models (LALMs)\citep{Audio-Flamingo,Qwen2-Audio,StepAudio2,mimoaudio} and Omni-Language Models (OLMs)\citep{qwen25omni,minicpmo,mingomni} has significantly advanced audio understanding. At the same time, it has spurred the development of numerous benchmarks to comprehensively evaluate their capabilities. Earlier benchmarks\citep{audiobench,airbench} mainly focused on semantic-level understanding tasks (transcription, captioning, and simple question answering), and recent benchmarks\citep{MMAU,ma2025mmar,mmaupro} have begun to investigate logical audio reasoning tasks. 

\begin{hlblock}
However, existing benchmarks largely overlook audio 4D intelligence. Although some advanced benchmarks do touch upon spatio-temporal aspects, their coverage remains limited in both scale and depth.  While MMAU \cite{MMAU}, MMAU‑Pro \cite{mmaupro} and MMAR \cite{ma2025mmar} contain temporal questions, they mainly involve identifying the timing or ordering of events  (e.g., when a sound occurs, which event comes first). These are primarily perceptual‑layer tasks. By contrast, our ``temporal deep reasoning'' tasks require understanding physical principles or causal dynamics across segments (e.g., inferring how a process evolves over time or how one event implies another), which cannot be solved by local timing cues alone. In addition, the spatial tasks in MMAR and MMAU-Pro are often restricted to single-source localization, and many items do not necessitate  meaningful use of stereo cues (e.g., simple arriving vs. departing judgments). In contrast, \ourbench introduces a hierarchical design covering three sub-tasks in complex scenes and explicitly emphasizes stereo-cue-based reasoning.

A comparative overview of \ourbench and prior benchmarks is presented in \cref{tab:benchmark_comparison}.
\ourbench evaluates deep spatio-temporal reasoning through tasks that go beyond surface‑level perception and instead require applying physical or causal knowledge, performing multi-step reasoning in complex real-world scenarios, and integrating information across multiple clips or events. \ourbench rests on a hierarchical and comprehensive task design. In addition, a rigorous data curation pipeline ensures high-quality samples, and robust evaluation strengthens the reliability.  
\end{hlblock}

\begin{table}[t]
\caption{A comparative overview of our benchmark against other representative audio benchmarks. 
(\cmark: Fully supported, \pmark: Partially supported or limitted amount, \xmark: Not supported)}

\label{tab:benchmark_comparison}
\vspace{-6pt}
\centering
\resizebox{\textwidth}{!}{%
\begin{tabular}{lccccccccc}
\hline
\multirow{2}{*}{\textbf{Benchmark}} & 
\makecell{\textbf{Temporal}\\\textbf{Deep Reasoning}} 
&\makecell{\textbf{Spatial}\\\textbf{Deep Reasoning}} 
& \makecell{\textbf{Quantitative }\\\textbf{Attribute}\\\textbf{Evaluation}} 
& \makecell{\textbf{Robust}\\\textbf{Evaluation}} 
& \makecell{\textbf{Multi-}\\\textbf{Audio}} 
& \makecell{\textbf{Fully}\\\textbf{Human-}\\\textbf{Annotated}} 
& \makecell{\textbf{ Fully  }\\\textbf{Expert}\\\textbf{Verified}} 
\\ \midrule
AIR-Bench [\citenum{airbench}] & \xmark & \xmark &\xmark &\xmark  &\xmark  & \xmark &\xmark  \\
MMAU [\citenum{MMAU}] & \xmark & \xmark &\xmark &\xmark  &\xmark  & \cmark &\cmark  \\
Dynamic-SUPERB Phase-2 [\citenum{dynamicphase2}] & \xmark & \xmark &\xmark &\xmark  &\pmark  & \pmark &\xmark  \\
MMAR [\citenum{ma2025mmar}]  & \xmark & \pmark &\xmark &\xmark  &\pmark  & \cmark &\cmark  \\ 
MMAU-Pro [\citenum{mmaupro}] &\xmark &\pmark &\xmark &\xmark &\cmark & \cmark & \cmark \\
\midrule
\textbf{\ourbench (ours)} & \cmark & \cmark & \cmark & \cmark & \cmark & \cmark & \cmark \\
\hline
\end{tabular}}
\vspace{-6pt}
\end{table}

%% file: 3-Method.tex
\section{\ourbench}\label{sec:benchmark}
\begin{figure*}[tb!]
\centering
\includegraphics[width=.98\columnwidth] {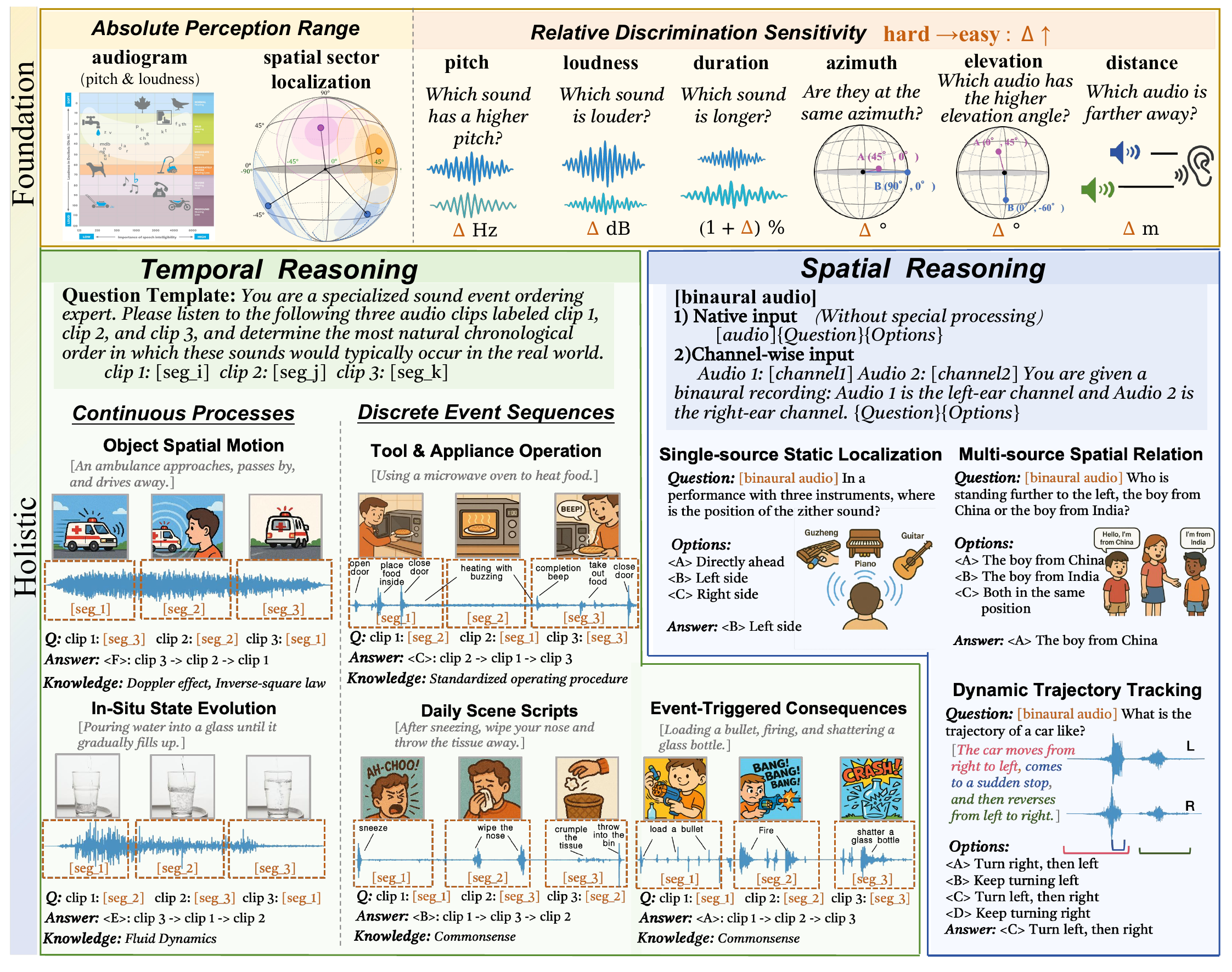} 
\vspace{-12pt}
\caption{\textbf{Data examples from \ourbench}: \textbf{(1)} the foundational perception task (upper) and \textbf{(2)} the holistic spatio-temporal reasoning task, which includes both temporal reasoning (bottom left) and spatial reasoning (bottom right). Zoom in for the best view.}
\label{fig:examples}
\vspace{-20pt}
\end{figure*}

Understanding dynamic sound sources in both time (1D) and three-dimensional space (3D) is a crucial skill for MLLMs to comprehend the physical world.
To address this need, our benchmark, \ourbench, is designed to comprehensively evaluate this 4D intelligence in the audio domain.
As illustrated in \cref{fig:examples}, our evaluation has two complementary sub-tasks: (1) Foundational Acoustic Perception (\cref{sec:foundational_perception}), which uses procedurally synthesized audio to quantitatively profile a model's basic perceptual abilities under controlled conditions, and (2) Holistic Spatio-Temporal Reasoning (\cref{sec:holistic_reasoning}), which uses real-world audio to evaluate more complex reasoning in dynamic and authentic scenarios.
We also elaborate our data curation pipeline in the \cref{sec:data_curation}.

\subsection{Foundational Acoustic Perception}
\label{sec:foundational_perception}
The Foundational Acoustic Perception task is motivated by the need for a robust, quantitative evaluation of the core perceptual abilities that underpin 4D audio intelligence.
A model's capacity for complex reasoning about dynamic audio scenes in the physical world is directly dependent on its ability to accurately perceive fundamental acoustic properties.
Our foundational acoustic perception task systematically probes a model's understanding of three critical auditory attributes: \textbf{Loudness}, \textbf{Pitch}, \textbf{Duration}, and the three spatial dimensions: \textbf{Azimuth}, \textbf{Elevation}, and \textbf{Distance}.
Just as a solid understanding of grammar is required for writing a complex narrative, a model must be able to accurately perceive these core attributes before it can reason about the dynamic, spatial relationships of sound sources in the physical world.
Without a firm grasp of these foundational elements, a model cannot accurately interpret complex, real-world acoustic scenes, which require understanding how sounds change over time and move through space.

We employ a targeted synthesis strategy to generate precise evaluation samples in a controlled environment for the foundational perception task.
For non-spatial attributes (Loudness, Pitch, Duration), we synthesize pure sine waves by directly specifying their parameters.
For spatial attributes (Azimuth, Elevation, Distance), we use the Pyroomacoustics \citep{scheibler2018pyroomacoustics} physics-based simulation engine to render acoustic scenes.
The targeted synthesis strategy allows us to investigate a model's audio perceptual abilities under the following two sub-tasks:

\textbf{1) Absolute Perception Range,} which defines the sensory limits of MLLMs for acoustic attributes.
For pitch and loudness, we adapt the design of human audiometry tests to create an ``audiogram'' for the MLLMs. Specifically, we synthesize sine waves with frequencies ranging from $125$ Hz to $8000$ Hz and loudness levels from $-10$ to $110$ dB HL and require the model to identify if a clear beep is in the first or second part of an audio clip, or if it's not there at all.
For spatial attributes, we design interval localization tasks that require the model to identify a sound's azimuth within one of four 90° quadrants (from 0° to 360°), its elevation relative to ear-level (above, at, or below, from -90° to 90°), and its distance category (near, medium, or far, within a 0 - 10m range). \cref{tab:fap_task_example} presents detailed examples of these absolute perception range tasks. Through these precise tasks, we establish the absolute limits of what the model can hear, which is crucial for developing AI systems that can safely and effectively interact with the physical world. 

\textbf{2) Relative Discrimination Sensitivity,} which investigates how well a model can detect small changes in acoustic attributes.
The ability to detect small changes allows a model to make nuanced judgments, like determining if a sound is getting louder or a pitch is rising.
Analogous to measuring the human Just Noticeable Difference (JND), the relative discrimination task presents the model with an audio clip containing two sounds and requires it to compare them based on a specific attribute.
We meticulously designed four to six distinct difficulty levels for each of the six attributes, as detailed in \cref{tab:fap_task_example}.
Level 1 serves as a control group to test for random guessing, presenting identical sounds ($\Delta$=0) for non-spatial attributes and a sub-threshold difference for spatial ones.  Subsequent levels then introduce progressively larger differences, ranging from subtle variations perceptible to humans to more significant, real-world changes.
By analyzing the model's performance across these different levels of stimulus differences, we can quantitatively assess its discrimination sensitivity for each attribute.

\subsection{Holistic Spatio-Temporal Reasoning}
\label{sec:holistic_reasoning}

Building on the model's fundamental audio perceptual abilities (\cref{sec:foundational_perception}), we further introduce holistic temporal reasoning (\cref{sec:temporal_reasoning}) and spatial reasoning (\cref{sec:spatial_reasoning}), which are designed to systematically evaluate a model's reasoning ability that is required for audio 4D intelligence.

\subsubsection{Temporal Reasoning Tasks}\label{sec:temporal_reasoning}
The core of temporal reasoning lies in understanding the intrinsic logic of event sequences, encompassing physical causality, functional procedures, or social conventions.
To evaluate this capability, we design a novel \textbf{Audio Segment Reordering} setting. 
Specifically, we curate a collection of audio events characterized by strong sequential uniqueness, semantic clarity, and logical universality.
Each event is segmented into three clips, which are then shuffled as inputs to the model.
The models are required to restore the original temporal sequence based solely on the audio content.
Our temporal reasoning tasks are organized into two meta-categories (continuous processes, discrete event sequences) and five subcategories based on their core logical principles.

The \textbf{continuous processes} assess a model's ability to track the subtle, continuous evolution of acoustic features within a single, uninterrupted acoustic event.
The \textbf{object spatial motion} subcategory reconstructs the spatio-temporal trajectory of moving sources (e.g., passing cars, airplanes) by interpreting key acoustic cues, such as the Doppler effect (frequency shifts indicating relative velocity) and the inverse-square law (loudness changes indicating distance).
Besides, the \textbf{in-situ state evolution} subcategory assesses a model's ability to track the intrinsic evolution of a stationary object's state, a process governed by predictable trend patterns.
These trend patterns arise from various underlying principles, including: \textit{Fluid \& Pneumatic Dynamics}, where the sound is governed by principles of turbulence, resonance, and pressure changes (e.g., a toilet flushing, water being poured); \noindent \textit{Thermodynamic Processes}, involving irreversible state changes driven by heat (e.g., water boiling, food frying); \textit{Energy Decay}, a process governed by resonant decay and frictional damping after a single excitation (e.g., a bell's chime, an explosion's echo); and complex \textit{Biological Rhythms} that reflect an evolving physiological or emotional state.

The \textbf{discrete event sequences} category requires the model to understand the logical and temporal relationships between multiple, distinct acoustic events, which are governed by function, convention, or causality.
The \textbf{tool \& appliance operation} sub-category follows the standardized operating procedure for tools and appliances (e.g., a microwave, a power drill), where the sequence is correct when it follows the tool's designed function.
The \textbf{daily scene scripts} sub-category applies commonsense and contextual script knowledge to follow the conventional sequence of actions in a daily activity (e.g., brushing teeth, drinking water).
The \textbf{event-triggered consequences} sub-category applies causal reasoning to infer that a trigger event (e.g., a firework explosion) will be followed by an automatic and irreversible outcome, whether physical (glass shattering) or social (a crowd cheering).

\subsubsection{Spatial Reasoning Tasks}\label{sec:spatial_reasoning}
Humans effortlessly perceive complex 3D auditory scenes (e.g., hearing a voice from behind, following an approaching car, or locating multiple speakers). Such an ability is fundamental for egocentric interaction and embodied AI systems, for instance, robots that navigate and interact with their surroundings. However, existing benchmarks focus primarily on the localization of static sound sources, whereas real-world scenarios demand reasoning that integrates both spatial and temporal cues. To address this gap, we organize the spatial reasoning task into three subcategories.

The \textbf{single-source static localization} evaluates the model’s ability to identify the direction of a target sound source among multiple static sources (e.g., judging whether a sound comes from the left or right). It assesses the basic spatial perception capability of the model and provides the foundation for more advanced reasoning.
The \textbf{multi-source spatial relation} requires the model to determine the relative spatial relationships among multiple simultaneous sound sources (e.g., comparing the placement of two speakers to decide which one is further to the right). Beyond localizing each source individually, the model must infer their spatial placement and choose the appropriate relational description from multiple candidates.
The \textbf{dynamic trajectory tracking} introduces moving sound sources, which require the model to go beyond basic spatial perception to dynamically model spatio-temporal relations for reasoning about complex movement trajectories (e.g., tracking a passing car moving from left to right). This task extends spatial reasoning into the temporal domain and is more faithful to the complexity of real-world acoustic scenarios.

\begin{wrapfigure}{r}{0.58\linewidth}
    \centering
    \vspace{-1.5em}
    \includegraphics[width=\linewidth]{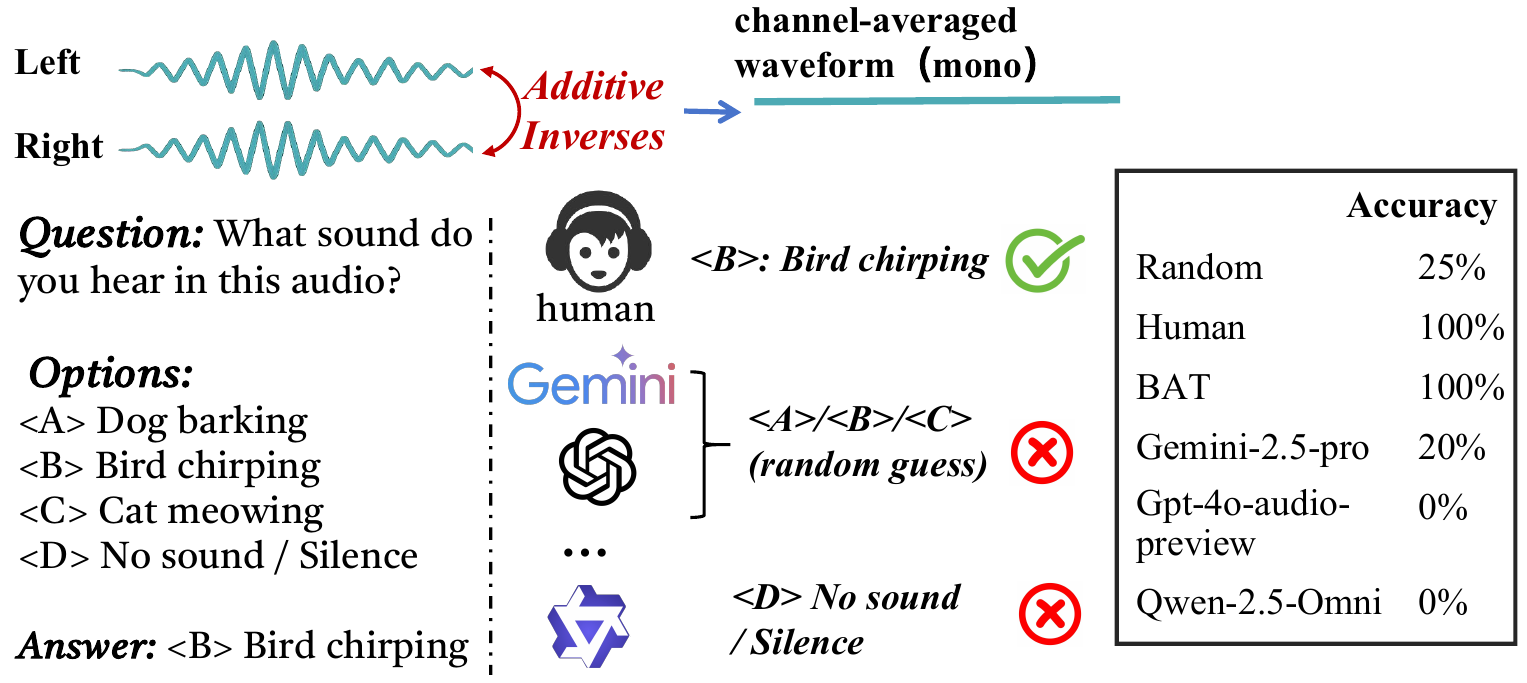}
    \vspace{-6pt}
    \caption{Audio preprocessing in existing models results in the loss of dual-channel information.}
    \label{fig:reverse}
    \vspace{-12pt}
\end{wrapfigure}

However, evaluating existing LALMs on multi-channel spatial tasks is challenging. The common practice of these models is to average multi-channel audio into a mono signal, resulting in the loss of substantial spatial information.
We conduct a simple experiment as shown in \cref{fig:reverse}.
We construct 20 pseudo-stereo signals by assigning the original audio to the left channel and its additive inverse to the right.
While human listeners could easily perform sound event classification on these signals, the models consistently failed due to signal cancellation during the mono conversion.
The result confirms their lack of explicit support for genuine stereo audio processing.
\begin{hlblock}
To provide a comprehensive assessment, we adopt two complementary strategies.
The first is the \textbf{native input} setting, where the model directly processes stereo audio using its default pipeline. This allows us to probe its intrinsic ability to exploit spatial cues.
The second is the \textbf{channel-wise input} setting, where the left and right channels are presented separately with explicit textual instructions, as shown in the bottom right of \cref{fig:examples}.
This configuration serves as an ablation study to examine whether current models have any spatial capability when the binaural information is preserved at the input.
\end{hlblock}


\subsection{Data Curation Pipeline}\label{sec:data_curation}
\begin{figure*}[t]
  \centering
  \begin{minipage}{0.7\textwidth}
    \centering
    \includegraphics[width=\linewidth]{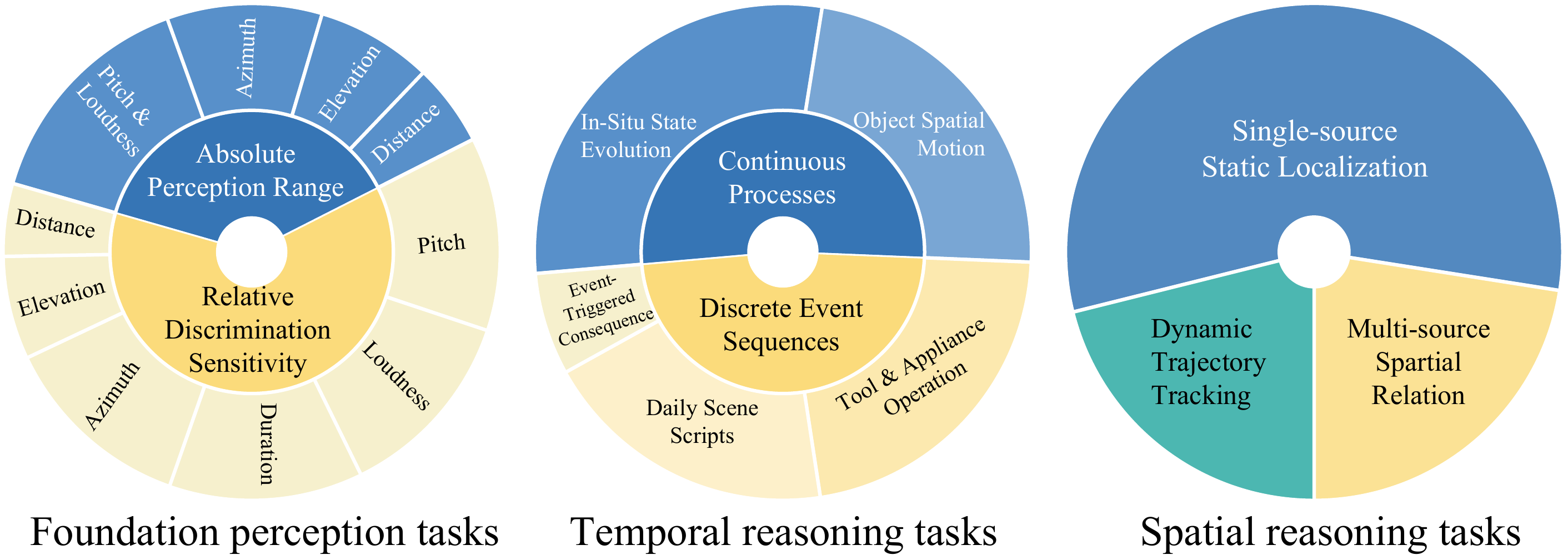}
    \subcaption{Data distribution across the foundation perception, temporal reasoning, and spatial reasoning three tasks.}
    \label{fig:modality_pie}
  \end{minipage}
  \begin{minipage}{0.28\textwidth}
    \centering
    \small
    \setlength{\tabcolsep}{.5pt}
    \resizebox{.75\textwidth}{!}{
    \begin{tabular}{lc}
      \toprule
      \textbf{Statistics} & \textbf{Number} \\
      \midrule
      Total Questions & 2{,}353 \\
      \midrule
      \textcolor{gray}{\textit{Foundation Perception}} \\
      Questions  & 951\\
      Category  & 2 \\
      Attributes  & 6 \\
      \midrule
      \textcolor{gray}{\textit{Temporal Reasoning}} \\
      Questions & 900 \\
      Category & 2 \\
      \textcolor{gray}{\textit{Spatial Reasoning}} \\
      Questions & 502 \\
      Category & 3 \\
      Avg. Audio Length  & 14{.}03 sec \\
      \bottomrule
    \end{tabular}}
    \subcaption{Statistics.}
    \label{tab:statistics}
  \end{minipage}
  \vspace{-6pt}
  \caption{\textbf{(a)} The \textbf{data distribution} of \ourbench across three main tasks. \textbf{(b)} \textbf{Data statistics} of our benchmark, including the total number of questions for each task and their sub-categories, and the average audio length for reasoning tasks.}
  \label{fig:combined_stat}
  \vspace{-12pt}
\end{figure*}

\begin{figure*}[tb!]
	\centering
	\includegraphics[width=0.9\columnwidth]{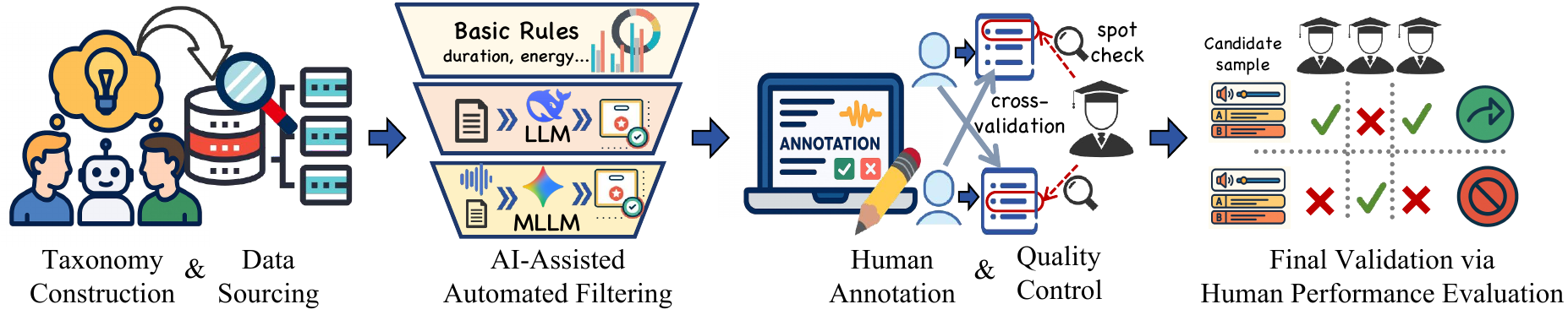}
    \vspace{-6pt}
	\caption{The four-stage \textbf{data annotation pipeline} for constructing our \ourbench.}
	\label{fig:data_pipeline}
    \vspace{-12pt}
\end{figure*}

Our data curation pipeline integrates procedural synthesis with real-world data collection to ensure both comprehensive coverage and ecological validity.
\cref{fig:combined_stat} shows the distribution and statistics of our \ourbench.
All audio for the \textit{foundational perception} task is synthesized using precise parameterization or the Pyroomacoustics \citep{scheibler2018pyroomacoustics} physics-based simulator, providing complete control over acoustic parameters.
Domain experts rigorously validate the task difficulty levels, which are then calibrated through human testing.
For the \textit{holistic spatio-temporal reasoning} task, the curation process comprises four key stages (see \cref{fig:data_pipeline}):

\textbf{1)} Taxonomy Construction and Data Sourcing: We build a hierarchical task taxonomy through a collaborative process involving domain experts and the Gemini 2.5 Pro \citep{gemini25pro}.
This framework guides the sourcing of candidate data from large-scale, real-world audio libraries: Clotho \citep{drossos2019clotho} and FSD50K \citep{fonseca2022FSD50K} for temporal reasoning, and STARSS23 \citep{shimada2023starss23}, along with audio sourced from the internet for spatial reasoning.

\textbf{2)} AI-Assisted Automated Filtering: This process employs an efficient three-stage funnel. First, we discard unsuitable samples based on basic properties like duration and energy. Next, an LLM (e.g., DeepSeek-V3 \citep{liu2024deepseekv3}) performs an initial screening based on textual metadata, providing justifications for its decisions. Finally, a powerful multimodal model (e.g., Gemini 2.5 Pro \citep{gemini25pro}) analyzes the audio, metadata, and the LLM's outputs.
The final step yields a judgment, a quality score, and a preliminary classification, further filtering irrelevant samples.
The detailed prompts used to query the LLMs are provided in \cref{appendix:filter_prompt}.

\textbf{3)} Human Annotation and Quality Control: We recruit and train 10 undergraduate annotators to label the data using a professional platform.
During this process, AI-generated information is provided as an auxiliary reference. To ensure high-quality labels, we implement a stringent two-round review process: the first round involves inter-annotator cross-validation until a consensus is reached, while the second consists of random spot-checks by three domain experts.
\begin{hlblock}
More details are provided in \cref{appendix:human_anno}.
\end{hlblock}

\textbf{4)} Final Validation via Human Performance Evaluation: To ensure all items in the benchmark are fair, unambiguous, and solvable by humans, we implement a final validation stage. In this phase, domain experts act as examinees and solve our tasks. Only items that are independently and correctly solved by at least two-thirds of the experts are retained. Our rigorous protocol ensures that all problems in our benchmark are well-posed and reliably solvable by human experts.

%% file: 4-Experiments.tex
\definecolor{acrgray}{gray}{0.55} 
\newcommand{\aaacr}[2]{\mbox{#1\,{\textcolor{acrgray}{/\,#2}}}} 
\newcommand{\na}{\textemdash} 

\section{Evaluation}
\noindent \textbf{Benchmarking Models.}
Our evaluation covers 19 models (16 open-source and 3 closed-source models). The open-source models span three categories: (1) Large Audio Language Models designed for universal audio-text understanding, including SALMONN \citep{tang2024salmonn}, Qwen2-Audio Instruct \citep{Qwen2-Audio}, Audio Flamingo 3 \citep{Audio-Flamingo3} with its `think' variant, DeSTA2.5-Audio \citep{lu2025desta2}, Kimi-Audio \citep{Kimi-Audio}, Step-Audio-2-mini \citep{StepAudio2}, MidashengLM \citep{midashenglm}, and Xiaomi-MiMo-Audio \citep{mimoaudio} with its `think' variant; (2) a specialized model for spatial audio, BAT \citep{zheng2024bat}; and (3) Omni Language Models with fully multimodal support, including Qwen-2.5-Omni \citep{qwen25omni}, Phi4-MM \citep{phi4mm}, Gemma-3n-E4B-it \citep{gemma3}, and Ming-Lite-Omni-1.5 \citep{mingomni}.
We also include three leading closed-source models: Gemini 2.5 Pro \citep{gemini25pro} (updated June 2025), Gemini 2.5 Flash (updated June 2025), and GPT-4o-audio-preview \citep{gpt4o} (version 2025-06-03).

\noindent \textbf{Robust Evaluation.}
All questions in \ourbench are presented as multiple-choice questions and evaluated using classification accuracy, with correctness determined via string matching of option labels or their full text. To ensure robustness, we evaluate each question multiple times under minor prompt perturbations, a strategy detailed in \cref{appendix:robust_eval}. This approach yields two key metrics: \textbf{Average Accuracy (AA)}, the mean accuracy across all runs, and \textbf{All-Correct Rate (ACR)}, the proportion of questions answered correctly in every run, which serves as a stronger indicator of model reliability. Due to space limitations, we primarily report AA in the main text, while complete experimental results are available in \cref{appendix:detail_res}.

\begin{table}[t]
\centering
\caption{%
Evaluation results of various models on \ourbench. The best performance is highlighted in \textbf{bold}, and the second-best ones are \underline{underlined}.
MA (Macro Accuracy) denotes the unweighted mean of class-wise accuracies, while OA (Overall Accuracy) denotes the proportion of correctly answered instances.
All reported values are AA (Average Accuracy across multiple runs) only; for ACR (All-Correct Rate), see \cref{appendix:detail_res}.}
\vspace{-6pt}
\label{tab:main_results}
\resizebox{\linewidth}{!}{%
\begin{tabular}{lcccccccccccc}
\toprule
\multirow{2}{*}{\textbf{Models}} & \multirow{2}{*}{\textbf{Size}} &\multicolumn{3}{c}{\textbf{Foundational Perception }} &
\multicolumn{3}{c}{\textbf{Temporal Reasoning }} &
\multicolumn{4}{c}{\textbf{Spatial Reasoning}} &
\multirow{2}{*}{\textbf{MA (\%)}} \\
\cmidrule(lr){3-5}\cmidrule(lr){6-8} \cmidrule(lr){9-12} 
& & \textbf{Range} & \textbf{Sensitivity} & \textbf{MA} & \textbf{Continuous } & \textbf{Discrete} & \textbf{OA} &
\textbf{Localization} & \textbf{Relation} & \textbf{Trajectory} & \textbf{OA}  & \\
\midrule
Random Guess & -  & 23.75 & 26.38 & 25.33 &14.29 &14.29 &14.29&33.33&33.33&33.33&33.33 &24.32\\
Human &- & 79.42 & 74.55 & 75.60 &90.12 &85.51 &88.00&70.00&80.00&77.00&73.72&79.11 \\
\midrule

SALMONN [\citenum{tang2024salmonn}]                  & 13B & 27.32 & 25.48 & 26.22 & 14.88 & 13.30 & 14.15 & 26.15 & 28.61 & 39.94 & 29.62 & 23.33\\
Audio Flamingo 3 [\citenum{Audio-Flamingo3}]        & 8.4B  & 31.79 & 35.72 & 34.15 & 9.23 &8.01 &8.67 & 37.22 & 38.35 & 44.03 & 38.91 & 27.24 \\
Audio Flamingo 3 think [\citenum{Audio-Flamingo3}] & 8.4B  & 25.54 & 34.08 & 30.66 &13.22 &14.02 &13.59 & 35.45 & 37.46 & 38.05 & 36.45 & 26.90 \\
Qwen2-Audio-Instruct [\citenum{Qwen2-Audio}]     & 8.4B  & 29.88 & 26.47 & 27.84 &13.29 &12.10 &12.74 &21.32 & 24.78 & 15.09 & 20.78 & 20.45 \\
DeSTA2.5-Audio [\citenum{lu2025desta2}] & 8.8B  & 29.87 & 19.79 & 23.82 &16.53 &17.39 &16.93 & 23.67 & 34.81 & 37.74 & 29.15 & 23.30 \\
BAT [\citenum{zheng2024bat}]  &7B & 22.81 & 6.25& 12.87 &0.00 &0.00 &0.00 & 0.00 & 0.00 & 0.00 & 0.00 & 4.29\\
Phi4-MM [\citenum{phi4mm}]  & 5.5B  & 19.14 & 29.85 & 25.56 &16.74 &16.99 &16.85 & 33.10 & 27.14 & 34.28 & 32.01 & 24.81 \\
Kimi-Audio [\citenum{Kimi-Audio}] & 7B & 23.29 & 27.50 & 25.82 &19.97 &16.83 &18.52&27.56&38.94&44.03&33.60 & 25.98 \\
MiDashengLM [\citenum{midashenglm}] &7B & \underline{36.94} & 30.78 & 33.24 &15.43 &17.31 &16.30&\textbf{43.11}&\underline{45.43}&\textbf{46.23}&\textbf{44.29} & 31.28 \\
Step-Audio-2-mini [\citenum{StepAudio2}]& 7B   & 29.65 & 27.14 & 28.14 &15.36 &15.87 &15.59&33.33&31.27&37.74&33.80 & 25.84 \\
Gemma-3n-E4B-it [\citenum{gemma3}]&7.5B  & 18.55 & 25.02 & 22.43 &16.87 &16.27 &16.59&23.32&41.89&33.96&29.75 & 22.92 \\
Ming-Lite-Omni-1.5 [\citenum{mingomni}]       & 18.9B   & 26.76 & 26.76 & 26.76 &17.08  &15.54 &16.37&20.14&35.10&38.36&27.35 & 23.49\\
Qwen-2.5-Omni [\citenum{qwen25omni}]           & 7B   & 28.76 & 32.32 & 30.90 &16.32 &17.71 &16.96&39.46&41.30&27.04&37.25 & 28.37 \\
Xiaomi-MiMo-Audio [\citenum{mimoaudio}]& 7B  & 34.95 & 31.59 & 32.93 &18.18 &19.15 &18.63&36.16&41.30&45.28&39.24 & 30.27 \\
Xiaomi-MiMo-Audio-think [\citenum{mimoaudio}]& 7B  & 29.90 & 24.93 & 26.92 &16.80 &19.39 &18.00&34.28&44.54&36.79&37.12 & 27.35 \\
MiniCPM-O-v2.6 [\citenum{minicpmo}] & 8B  & 31.02 & 31.87 & 31.53 &15.36 &17.39 &16.30&29.92&43.36&38.36&34.73 &27.52 \\
\cdashline{1-13}
\noalign{\vskip 0.4mm}
GPT-4o Audio [\citenum{gpt4o}]      &- & 27.58 & 34.55 & 31.76 &15.91  &23.56  &19.44&\underline{41.81}&43.97&39.94&41.70 & 30.97 \\
Gemini 2.5 Flash [\citenum{gemini25pro}]  &- & 33.46 & \underline{43.88} & \underline{39.72} &\underline{27.55} &\underline{34.38}  & \underline{30.70} & 24.62&43.07&22.64&28.35 & \underline{32.92}   \\
Gemini 2.5 Pro [\citenum{gemini25pro}]    &- & \textbf{39.90} & \textbf{51.13} & \textbf{46.64} &\textbf{54.88}  &\textbf{62.74}  &\textbf{58.52} &40.87&\textbf{48.97}&\underline{45.28}&\underline{43.62} & \textbf{49.59} \\
\bottomrule
\end{tabular}}
\vspace{-12pt}
\end{table}

\begin{figure*}[tb!]
	\centering
	\includegraphics[width=\columnwidth]{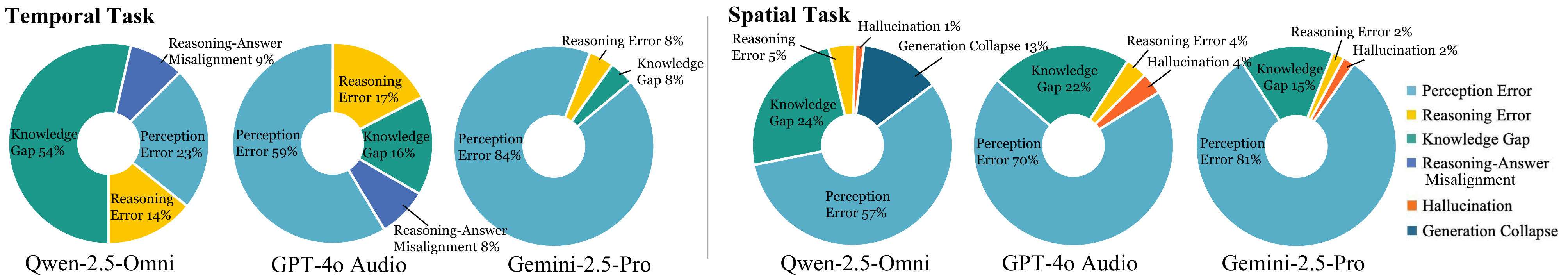}
    \vspace{-12pt}
	\caption{Error distribution across temporal and spatial Tasks.}
	\label{fig:error_distribution}
    \vspace{-12pt}
\end{figure*}

\subsection{Main Result Analysis}
We present a comprehensive evaluation on \ourbench, as shown in \cref{tab:main_results}. Due to the space limit, detailed results on each task are provided in \cref{appendix:detail_res}. Our key findings are as follows:

\textbf{\ourbench is Challenging}
\ourbench presents a considerable challenge for existing models. Human evaluators achieve high accuracy across all task categories (e.g., 75.6\% on perception, 88.0\% on temporal, and 73.7\% on spatial tasks), whereas all tested models fall well below this baseline. Most open-source models perform close to random guessing, and even the best closed-source model, Gemini 2.5 Pro, reaches only 49.59\% average accuracy.
In addition, model predictions on \ourbench exhibit low reliability, as evidenced by the pronounced gap between their Average Accuracy (AA) and All-Correct-Rate (ACR) scores. A detailed discussion of this issue is provided in \cref{appendix:instability}.
Although the underlying audio data for the temporal tasks (e.g., FSD50K, Clotho) is commonly used for model pre-training, our novel task formulation of temporal reasoning deliberately departs from conventional audio QA formats. This design allows for a more thorough evaluation of the integrated capabilities of current models.
\begin{hlblock}
Meanwhile, this design also serves as a diagnostic lens on the limitations of current training pipelines.
The poor performance across models suggests that existing training paradigms often centered on clip‑level tagging, QA, or captioning over linguistically salient cues (e.g., using FSD50K for sound event recognition) and do not equip models with the abilities needed for audio 4D intelligence.
\end{hlblock}

\textbf{A Clear Performance Gap between Closed-Source and Open-Source Models}
On the foundational perception and temporal tasks, Gemini 2.5 Pro establishes a commanding lead among all models. On spatial tasks, however, nearly all models, both closed- and open-source, perform poorly. As indicated by the prior experiment (\cref{fig:reverse}), this is likely because most models (except BAT) discard multi-channel information during preprocessing, thereby losing key acoustic cues needed for spatial reasoning.
Among closed-source models, Gemini 2.5 Pro surpasses Gemini 2.5 Flash, suggesting that stronger reasoning capabilities deliver substantial gains. In contrast, open-source models show the opposite pattern: the ``think'' modes of Audio Flamingo 3 and Xiaomi-MiMo-Audio perform worse than their no-thinking counterparts, implying that without sufficiently solid perceptual and knowledge foundations, reasoning can be ineffective or even detrimental.

\subsection{Discussion: Why Do Existing Models Struggle on \ourbench?}
To better understand the underlying causes of the poor performance of existing models, we conduct a detailed error analysis along with a series of ablation studies. Due to space limitation, the ablation study on spatial reasoning is provided in \cref{appendix:spatial_ablation}.

\begin{figure*}[tb!]
	\centering
	\includegraphics[width=0.95\columnwidth]{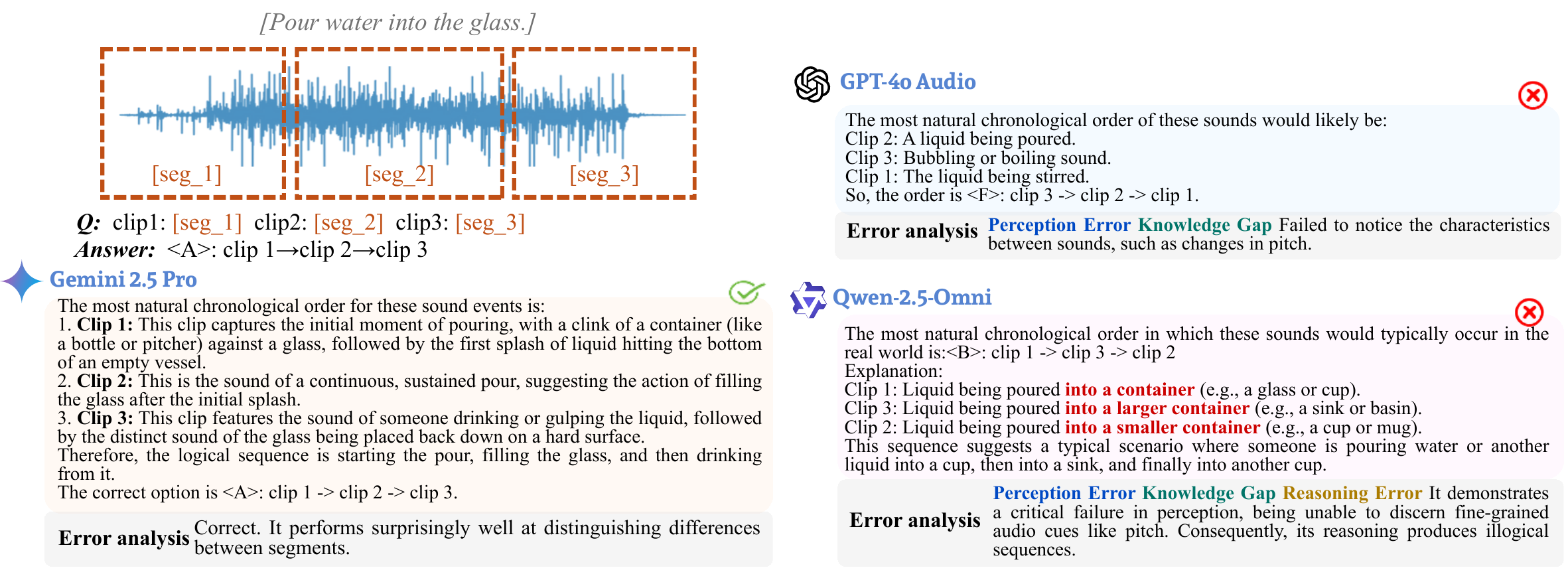}
    \vspace{-12pt}
	\caption{An error case in temporal reasoning task. More cases are provided in the \cref{appendix:cases}.}
	\label{fig:temporal_case}
    \vspace{-12pt}
\end{figure*}

\noindent \textbf{Error Analysis.}
We conduct a manual error analysis on 200 failed predictions sampled equally from temporal and spatial tasks of three representative models (Gemini 2.5 Pro, GPT-4o-audio, and Qwen-2.5-Omni).
For temporal tasks, our analysis reveals a clear capability hierarchy across the models. The open-source Qwen-2.5-Omni shows major deficiencies in all three core abilities: its perception is coarse-grained and unable to capture subtle inter-segment distinctions, and a substantial knowledge gap (54\%) leads to reasoning that often appears specious due to the absence of physical-world grounding. GPT-4o-audio demonstrates stronger knowledge, but still suffers from perceptual and reasoning limitations, along with low-level issues such as misalignment between reasoning and final answers. In contrast, Gemini 2.5 Pro excels in knowledge and reasoning, shifting its primary bottleneck to the more advanced challenge of fine-grained perception (84\%). As shown in \cref{fig:temporal_case}, Gemini 2.5 Pro is the only model to succeed by providing a remarkably detailed description of acoustic nuances. Our finding suggests that the \textbf{advanced world knowledge is deeply embedded within detailed audio-text captioning.} While open-source models largely remain at a coarse semantic level (e.g., sound event classification), our analysis highlights that enabling them to generate fine-grained acoustic descriptions is critical toward more robust reasoning.
On the other hand, most models demonstrate a lack of native spatial awareness in audio tasks, with weaknesses in perception, knowledge, and reasoning. Additionally, a prevalent type of error involves vision-centric hallucinations (e.g., ``...based on the car's trajectory in the video...''). This may be attributable to the models' training on visual spatial tasks, leading them to misapply visual reasoning to auditory inputs.

\begin{figure*}[t]
\centering
\begin{minipage}{.56\textwidth}
\includegraphics[width=\linewidth]{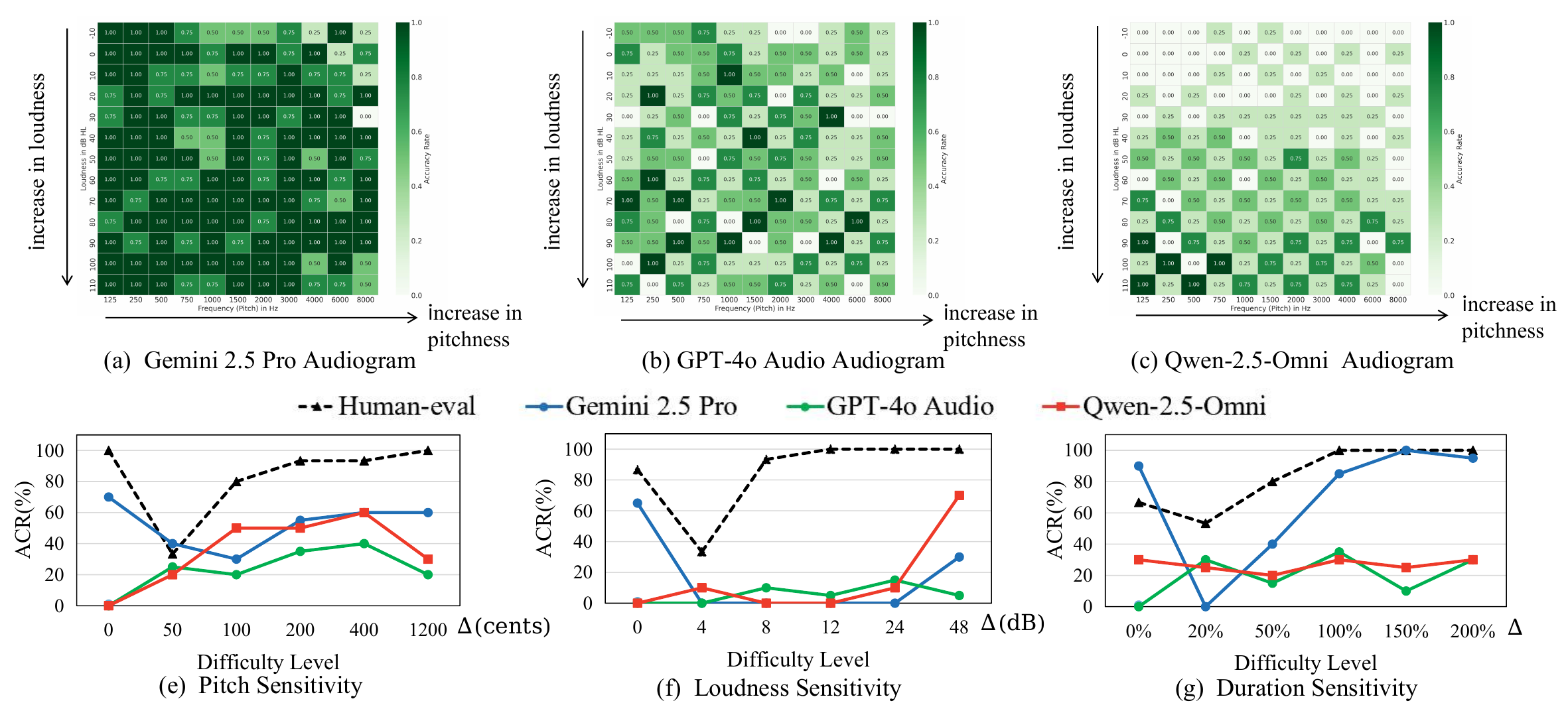}
\vspace{-12pt}
\caption{\begin{hlblock}
   The \textbf{range} and \textbf{sensitivity analysis} in foundational perception. 
\end{hlblock}}
\vspace{-12pt}
\label{fig:gram_curve}
\end{minipage}
\hspace{-1pt}
\begin{minipage}{.43\textwidth}
\vspace{6pt}
\includegraphics[width=\linewidth]{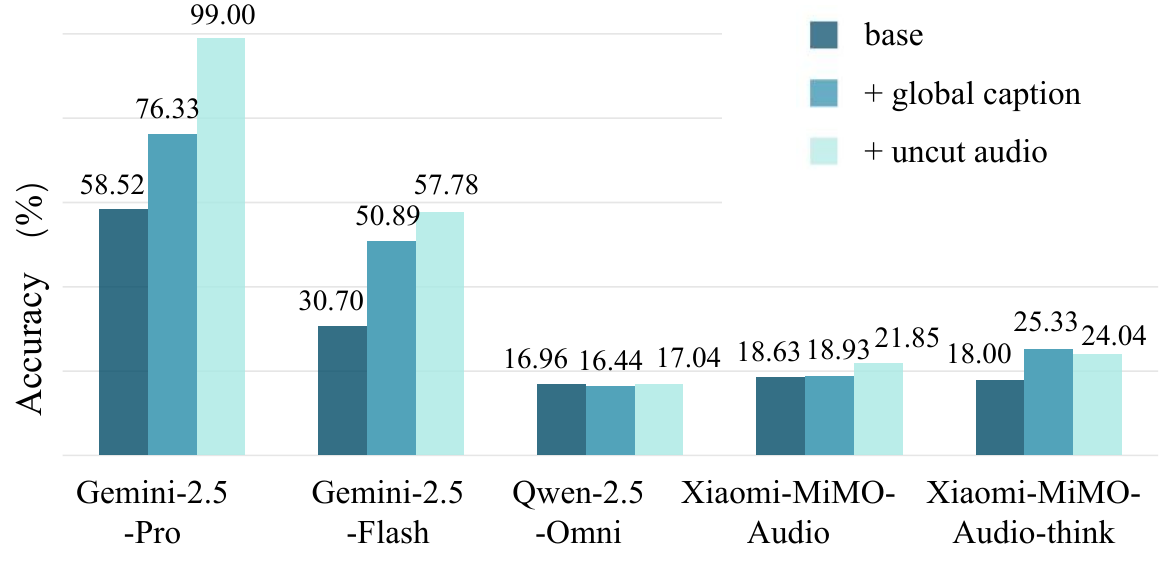}
\vspace{-8pt}
\caption{The \textbf{ablation study} on temporal reasoning.}
\label{fig:temporal_ablation}
\vspace{-12pt}
\end{minipage}
\end{figure*}

\begin{hlblock}
\noindent \textbf{Lack of Human-like Range and Sensitivity in Foundational Perception.}
To quantify the gap in perceptual range and sensitivity, we provide detailed visualizations of model performance on our foundational perception tasks in \cref{fig:gram_curve}.
The first row of \cref{fig:gram_curve} presents audiograms that compare model coverage across the pitch–loudness space. Gemini 2.5 Pro achieves a much broader coverage than the other two models, where greener regions indicate higher accuracy and the covered area reflects the perceptual range. In contrast, human listeners with normal hearing are expected to achieve near-full coverage, underscoring the gap between current models and human perceptual abilities in terms of range.
\end{hlblock}
The second row of \cref{fig:gram_curve} further track the performance of both models and human subjects on the three core acoustic attributes (pitch, loudness, and duration) as task difficulty decreases. The results reveal a stark performance gap between all models and the human baseline, particularly in the perception of fine-grained loudness differences.
A clear trend is visible even for the top-performing Gemini 2.5 Pro: its accuracy, while competent on easier tasks, plummets as perceptual granularity increases. This directly corroborates our error analysis, identifying fine-grained perception as its primary bottleneck. Notably, its performance on duration perception is an exception, showcasing \textbf{temporal grounding capabilities superior to those of other models} by accurately assessing audio segment lengths.

\noindent \textbf{Ablation Study on Temporal Reasoning.}
To further pinpoint the specific limitations of temporal reasoning, we augment the baseline audio segment reordering task with two progressively easier settings: (1) \textit{+ Global Caption}, where a single sentence describing the overall scene is provided as a contextual guide; and (2) \textit{+ Uncut Audio}, where the complete, unsegmented audio track is offered as a reference, reducing the task to a straightforward process where the correct order can be determined simply by comparing and grounding each segment within the full audio.
As shown in \cref{fig:temporal_ablation}, Gemini 2.5 Pro’s performance scales effectively with task simplification, culminating in a near-perfect 99\% accuracy in the \textit{+ Uncut Audio} setting. In contrast, the open-source models show minimal to no improvement across these settings. Their performance remains stagnant even when provided with the complete audio reference, despite the simplified nature of the task. This finding starkly exposes a core weakness in current open-source models: \textbf{a fundamental inability to effectively compare, ground, and integrate information from multiple audio inputs.}

%% file: 5-Conclusion.tex
\section{Conclusion}
We introduce \ourbench, a comprehensive benchmark for evaluating 4D audio intelligence over time and 3D space.
We use rigorous human annotation, consensus review, and expert validation to ensure the high quality of data samples.
\ourbench establishes standardized tasks and protocols for studying 4D audio intelligence, offering actionable diagnostics for model developers.
We expect STAR-Bench to accelerate progress on advanced audio models and training with spatialized corpora, capabilities that are crucial for embodied agents.

%% file: 6-Appendix.tex
\newpage
\section*{The Use of Large Language Models}
We used Gemini-2.5-Pro to assist in expanding and consolidating the taxonomy of tasks in our benchmark. Both DeepSeek-V3 and Gemini-2.5-Pro were utilized for the automated pre-screening of candidate data.
The final task definitions and data samples are verified by humans.
We also used GPT-4o to generate some of the illustrative figures presented in the paper, and used GPT-5 to polish the manuscript text.
Only human-verified revisions are included in the final version.

\section{Related Work}
\label[appendix]{appendix:related}
\subsection{Audio Language Models}
With the advancements of large language models (LLMs) and multimodal language models \citep{qwen3,mixtral,gpt4o,gemini25pro,internlm,llama,liu2024miadpo, liu2025visualagentic, zhang2025sec, qi2025vlnr1,xing2025scalecap,xing2025pyramiddrop, ding2025mmifengine,wei2025simcot,wei2025videorope,li2025fixed, zhang2025booststep}, recent research has increasingly focused on integrating audio perception with LLMs to enhance audio understanding and reasoning. Existing methods can be broadly grouped into two categories: Large Audio Language Models(LALMs) and Omni Language Models(OLMs). 

Most LALMs combine a pre-trained audio encoder with an LLM backbone, where the two modalities are aligned via large-scale text-audio joint training. Notable models include LTU-AS  \citep{LTUAS}, SALMONN  \citep{tang2024salmonn}, MidashengLM  \citep{midashenglm}, Audio Flamingo series \citep{Audio-Flamingo2,Audio-Flamingo3}, Qwen-Audio series \citep{Qwen-Audio,Qwen2-Audio}, Step-Audio \citep{StepAudio2} and Mimo-Audio \citep{mimoaudio}. These models have achieved remarkable performance across a wide range of audio understanding tasks, including automatic speech recognition(ASR), spoken question answering(SpokenQA), and automated audio captioning(AAC). In parallel, OLMs extend this paradigm to unify multimodal understanding with representative examples such as Qwen-2.5-Omni \citep{qwen25omni}, Ming-Omni \citep{mingomni},MiniCPM-O \citep{minicpmo}, Phi-4 \citep{phi4mm}, GPT-4o \citep{gpt4o}, and Gemini 2.5 \citep{gemini25pro}.  Notably, they also achieve impressive performance on audio understanding and reasoning, highlighting their potential to bridge multimodal perception and advanced audio intelligence.

\subsection{Audio Benchmarks}

Existing audio benchmarks illustrate the rapid progress of multimodal evaluation but also expose limitations. AudioBench \citep{audiobench} and AIR-Bench \citep{airbench} primarily focus on tasks such as automatic speech recognition (ASR), spoken question answering (SpokenQA), and audio captioning (AAC). These settings tend to reduce audio understanding to transcription or description, thereby neglecting the broader spectrum of acoustic reasoning. MMAU \citep{MMAU} and MMAR \citep{ma2025mmar} further extend the evaluation scope. However, their results reveal an inherent weakness—LLMs equipped with audio captions can perform on par with advanced LALMs, suggesting that such benchmarks still probe little beyond language-level semantics. 

Although some advanced benchmarks, such as MMAR~\citep{ma2025mmar} and MMAU-Pro~\citep{mmaupro}, do touch upon spatio-temporal aspects, their coverage remains limited in both scale and depth. For instance, their temporal analysis is typically reduced to identifying the timing or order of events occurring in the audio, while spatial analysis is often limited to localizing a single sound source. In contrast, our benchmark systematically evaluates models’ temporal and spatial deep reasoning capabilities within complex, real-world physical contexts, requiring them to infer causal and dynamic relationships.
Beyond audio benchmarks, multimodal benchmarks in video question answering \citep{vstar,mmsi} and embodied AI \citep{embodiedbench} have emphasized temporal and spatial reasoning. However, these frameworks are predominantly grounded in the visual modality, where exploration of the audio modality remains comparatively limited. 
In real-world scenarios, audio understanding often depends on integrating information across multiple sound streams and reasoning about subtle changes in intensity, phase, or frequency—capabilities that existing benchmarks scarcely capture.

Our benchmark aims to address these gaps by introducing tasks that require \textbf{multi-audio input and cross-audio reasoning}, such as comparing or integrating information across multiple sound inputs, as well as \textbf{fine-grained spatio-temporal deep reasoning}, such as tracking how acoustic patterns evolve with underlying physical changes. Rather than being limited to surface-level semantics, the benchmark is designed to assess whether models can leverage raw audio cues to perform physically grounded reasoning across spatial and temporal dimensions.

\section{Details of Data Annotation}\label[appendix]{appendix:details_data}
In this section, we present the details of data annotation.

\subsection{Prompts for audio captioning}
\label[appendix]{appendix:caption_prompt}
The prompt for Gemini 2.5 Pro audio captioning: ``Please provide a detailed description of the audio, including speech, music, environmental sounds, and any other noticeable elements. Be as specific as possible.''

\subsection{Detail information for foundational acoustic perception}
\cref{tab:fap_task_example} details the ranges and levels used for each acoustic attribute, alongside illustrative examples of our foundational acoustic perception tasks. 

\begin{table*}[tp]
    \centering
    \setlength\tabcolsep{4pt}
    \resizebox{1.0\textwidth}{!}{
        \begin{tabular}{c|c|c}
        \toprule
        \textbf{Attribute}  &\textbf{Range / Level} &  \textbf{Example} \\
        \midrule
        \multicolumn{3}{c}{\textbf{Absolute Perception Range}} \\
        \midrule
        
        \multirow{3}{*}{\makecell{Pitch,\\ Loudness}}  & \multirow{3}{*}{\makecell{125 Hz - 8000 Hz\\-10dB - 110dB}}  & 
        \cellcolor{gray!5}{\textit{\darkblue{[Audio]The audio you just heard is divided into two halves. }}} \\
        ~ & ~  & \cellcolor{gray!5}{\textit{\darkblue{Does a sound appear in the first half, the second half, or is it not present at all?  }}} \\
        ~ & ~  & \cellcolor{gray!5}{(A) The first half   (B) The second half (C) It is not present at all (D) Unable to determine} \\
        \hhline{-|-|-} 
         \multirow{2}{*}{Azimuth} & \multirow{2}{*}{0° - 360°} & \cellcolor{gray!5}{\textit{\darkblue{[Audio] Given that 0° is directly in front and the angle increases clockwise, which azimuth range is the sound most likely coming from?}}} \\
        ~ & ~  & \cellcolor{gray!5}{(A) Front-Right (0°–90°) (B) Back-Right (90°–180°) (C) Back-Left (180°–270°) (D) Front-Left (270°–360°) (E) Unable to determine} \\
        \hhline{-|-|-}
         \multirow{2}{*}{Elevation} & \multirow{2}{*}{-90° - 90°} & \cellcolor{gray!5}{\textit{\darkblue{[Audio] Where does the sound seem to be coming from in terms of elevation, relative to ear level?}}} \\
        ~ & ~  & \cellcolor{gray!5}{(A) Above ear level (B) Below ear level (C) At ear level (D) Unable to determine}  \\
        \hhline{-|-|-}
         \multirow{2}{*}{Distance} & \multirow{2}{*}{0 meter - 10 meters} & \cellcolor{gray!5}{\textit{\darkblue{[Audio] How far away does the sound seem to be?}}} \\
        ~ & ~  & \cellcolor{gray!5}{(A) Near (within about 0–3 meters) (B) Medium (around 3–8 meters) (C) Far (more than 8 meters) (D) Unable to determine}  \\
        \midrule
        \multicolumn{3}{c}{\textbf{Relative Discrimination Sensitivity}} \\
        \midrule
        \multirow{2}{*}{Pitch}& \multirow{2}{*}{0, 50, 100, 200, 400, 1200 (cents)}  & \cellcolor{gray!5}{\textit{\darkblue{[Audio] Which sound has a higher pitch: the first sound, the second sound, or are they the same?}}} \\
        ~ & ~  & \cellcolor{gray!5}{(A) The first sound has a higher pitch (B) The second sound has a higher pitch(C) Both sounds are the same (D) Unable to determine} \\
        \hhline{-|-|-}
        \multirow{2}{*}{Loudness} & \multirow{2}{*}{0,  4, 8, 12, 24, 48 (dB)}  & \cellcolor{gray!5}{\textit{\darkblue{[Audio] Which sound is louder: the first sound, the second sound, or are they the same?}}} \\
        ~ & ~  & \cellcolor{gray!5}{(A) The first sound is louder (B) The second sound is louder (C) Both sounds are the same (D) Unable to determine} \\
        \hhline{-|-|-}
        \multirow{2}{*}{Duration} & \multirow{2}{*}{0, 20, 50, 100, 150, 200 (\%)}  & \cellcolor{gray!5}{\textit{\darkblue{[Audio] Which sound is longer: the first sound, the second sound, or are they the same?}}} \\
        ~ & ~  & \cellcolor{gray!5}{(A) The first sound is longer (B) The second sound is longer (C) Both sounds are the same (D) Unable to determine} \\
         \hhline{-|-|-}
         \multirow{2}{*}{Azimuth} & \multirow{2}{*}{30, 60, 90, 120, 150, 180 (°)} & \cellcolor{gray!5}{\textit{\darkblue{Audio 1: [Audio$\_$1]  Audio 2:[Audio$\_$2] Are Audio 1 and Audio 2 at the same azimuth? (Consider differences of less than 45° as the same.)}}} \\
        ~ & ~  & \cellcolor{gray!5}{(A) Same (B) Different (C) Unable to determine} \\
        \hhline{-|-|-}
         \multirow{2}{*}{Elevation} & \multirow{2}{*}{15, 90, 120, 150 (°)} & \cellcolor{gray!5}{\textit{\darkblue{Audio 1: [Audio$\_$1]  Audio 2:[Audio$\_$2] Which audio has the higher elevation angle? (Consider differences of less than 45° as the same.)}}} \\
        ~ & ~  & \cellcolor{gray!5}{(A) Audio 1 is higher (B) Audio 2 is higher (C) Both are at the same elevation (D) Unable to determine}  \\
        \hhline{-|-|-}
         \multirow{2}{*}{Distance} & \multirow{2}{*}{1-2, 4-5, 6-7, 8-9 (meters)} & \cellcolor{gray!5}{\textit{\darkblue{Audio 1: [Audio$\_$1]  Audio 2:[Audio$\_$2] Which audio is farther away? (Consider differences of less than 3 meters as the same.)}}} \\
        ~ & ~  & \cellcolor{gray!5}{(A) Audio 1 is farther away (B) Audio 2 is farther away (C) Both audios are the same (D) Unable to determine}  \\
        \midrule
        \end{tabular}
    }
    \caption{Task examples of foundational acoustic perception. 
   }
    \label{tab:fap_task_example}
    \vspace{-0.3cm}
\end{table*}

\subsubsection{Binaural Audio Synthesis}

We generated binaural recordings for foundational perception tasks (azimuth, elevation, distance) in Pyroomacoustics \citep{scheibler2018pyroomacoustics} across three rectangular rooms—small (4.0×3.5×2.8 m), medium (8.0×6.0×3.5 m), and large (20×15×8 m)—each with a frequency-independent wall absorption coefficient of 0.25. Image-source reflections were modeled up to order 10 at 44.1 kHz (matched to the HRTF sampling rate). For each room, we evaluated two listener positions (distinct Cartesian coordinates) and oriented the head toward the +x axis. Binaural reception used a co-located two-microphone array at the listener position with ear-specific directivity derived from a measured SOFA HRTF\footnote{\url{https://sofacoustics.org/data/database/mit/mit_kemar_normal_pinna.sofa}} (MIT KEMAR, “normal pinna”; interpolation order 12, 1000 points), loaded via a local SOFA reader and applied to the left/right channels.

For each condition (room × listener), sources were placed on a sphere centered at the listener (radii 1–10 m; configurable azimuth/elevation), and ear-specific BRIRs were computed. Mono source signals were drawn from three curated audio clips (“alarm,” “applause,” “telephones”), downmixed if necessary. Rendering was performed by convolving each dry signal with the left/right BRIRs after an early/late mix to emphasize distance cues: we preserved the first 80 ms and attenuated the late tail by 0.5. We then applied global peak normalization across the batch to avoid clipping while preserving inter-position level differences.

We discretized each attribute into fixed partitions to control dataset balance.

\textbf{Absolute azimuth:}
Eight angles
\(\{30^\circ,\,60^\circ,\,120^\circ,\,150^\circ,\,210^\circ,\,240^\circ,\,300^\circ,\,330^\circ\}\).
For each angle we rendered all combinations of 3 rooms \(\times\) 2 listener positions \(\times\) 2 source clips, yielding
\(8 \times (3 \times 2 \times 2) = 96\) utterances.
\textbf{Absolute elevation:}
Six angles
\(\{-75^\circ,\,-45^\circ,\,-15^\circ,\,15^\circ,\,45^\circ,\,75^\circ\}\).
Per angle we rendered 3 rooms \(\times\) 2 listener positions \(\times\) 2 source clips, for
\(6 \times (3 \times 2 \times 2) = 72\) utterances.
\textbf{Absolute distance:}
Radii from 1–10 m with a nonuniform allocation to emphasize near-field cues:
for 1–7 m we generated 6 utterances per meter (42 total),
and for 8–10 m we generated 3 per meter (9 total),
giving \(42+9=51\) utterances per (room \(\times\) listener) set.

\textbf{Relative azimuth:}
Differences were multiples of \(30^\circ\):
\(\{30^\circ,\,60^\circ,\,90^\circ,\,120^\circ,\,150^\circ,\,180^\circ\}\) (6 levels),
totaling \(6 \times 20 = 120\) utterances.
\textbf{Relative elevation:}
Four difference angles \(\{15^\circ,\,90^\circ,\,120^\circ,\,150^\circ\}\)
with 18, 17, 17, 12 utterances respectively (64 total).
\textbf{Relative distance:}
Four difference levels \(\{1-2,\,4-5,\,6-7,\,8-9\}\) m with counts per level
\(\{12,\,12,\,12,\,9\}\), totaling \(45\) utterances.

\begin{hlblock}
\subsection{Details of the Curation Process for Reasoning Tasks}
\end{hlblock}
\subsubsection{Prompt Used for AI-Assisted Filtering of Temporal Task Data}
\begin{figure*}[b!]
	\centering
	\includegraphics[width=0.95\columnwidth]{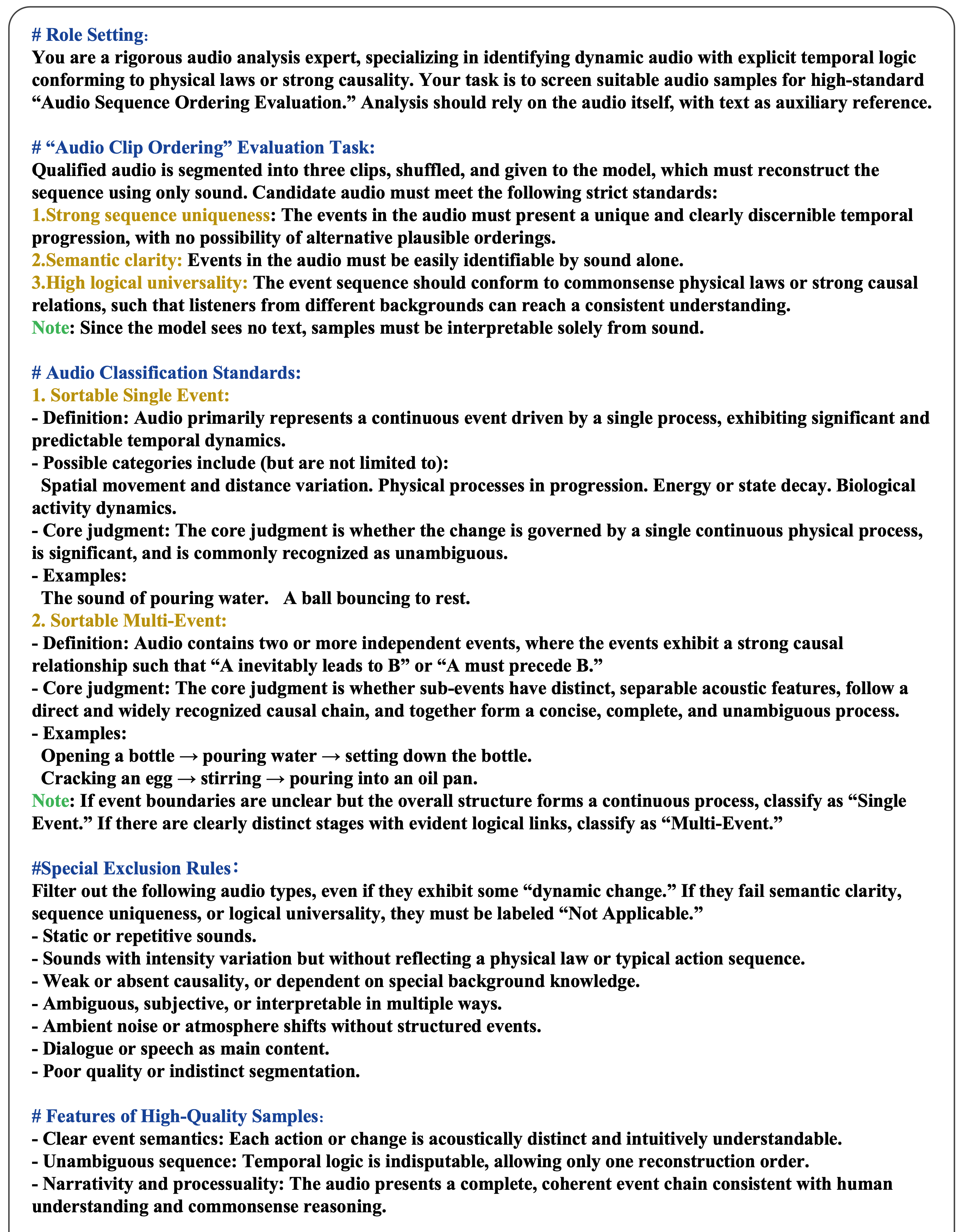}
    \vspace{-10pt}
	\caption{The prompt for our AI-assisted filtering process on temporal tasks.
    }
	\label{fig:duration_prompt1}
    \vspace{-10pt}
\end{figure*}

\begin{figure*}[tb!]
	\centering
	\includegraphics[width=0.95\columnwidth]{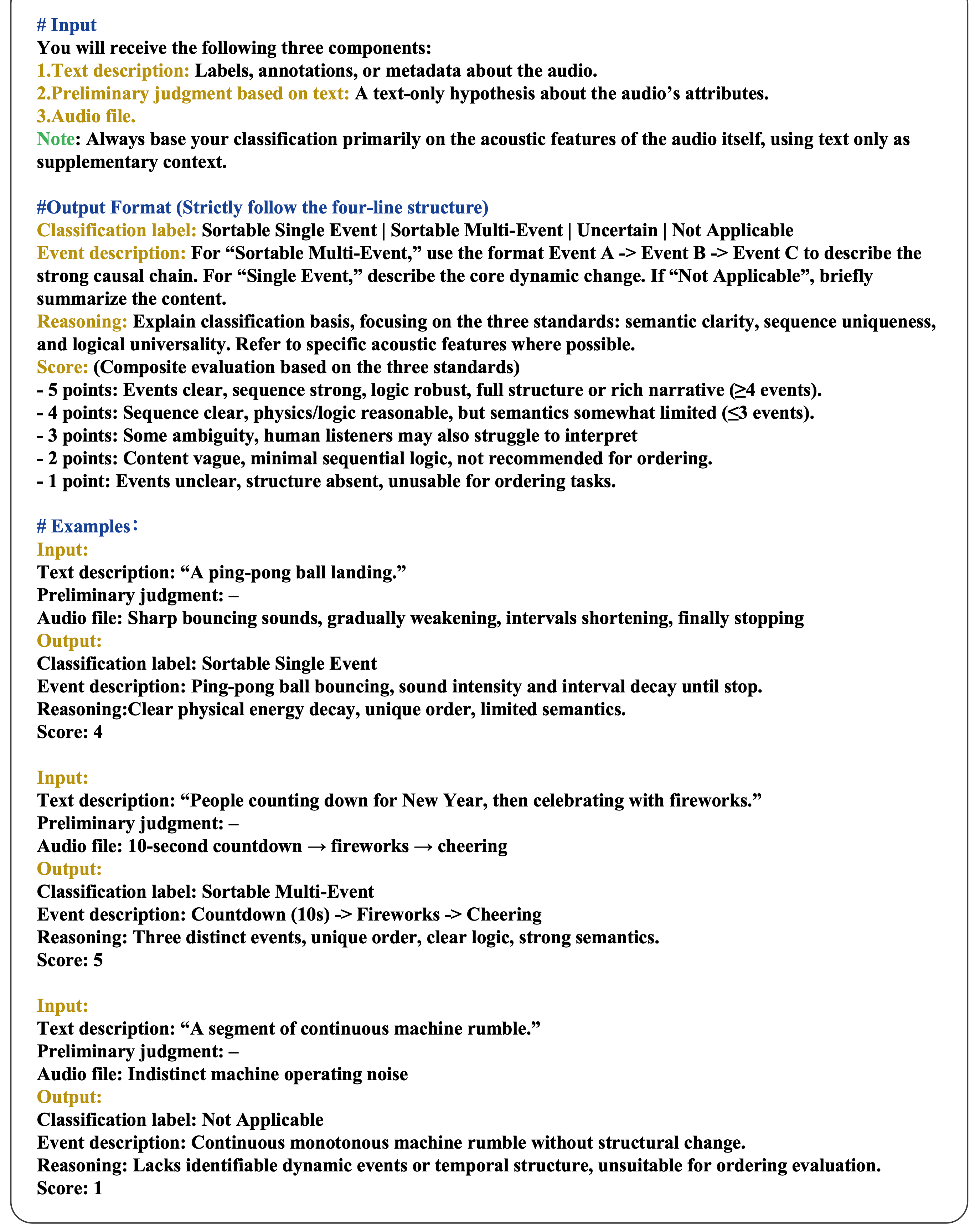}
    \vspace{-10pt}
	\caption{The prompt for our AI-assisted filtering process on temporal tasks.
    }
	\label{fig:duration_prompt2}
    \vspace{-10pt}
\end{figure*}

\label[appendix]{appendix:filter_prompt}
\cref{fig:duration_prompt1} and \cref{fig:duration_prompt2} present our carefully designed prompts, which leverage Gemini 2.5 Pro to filter candidate data that meet the requirements of audio segment reordering.
\begin{hlblock}
Briefly, we feed the audio, its metadata, and our task description, and ask Gemini 2.5 Pro to decide, under our strict criteria of strong sequence uniqueness, semantic clarity, and high logical universality, (i) whether the audio is suitable for a reordering task, (ii) whether it reflects a continuous or discrete process, (iii) the reasoning behind its judgment, and (iv) a quality score. We adopt a conservative filtering strategy, discarding only samples explicitly marked as ``not applicable''. All remaining clips, along with the model's analysis, are then passed to professional annotators for verification and annotation. A prior LLM-based filtering step follows a similar procedure, but without audio input. 
\end{hlblock}


\begin{hlblock}
\subsubsection{Details of Human Annotation and Quality Control}
\label[appendix]{appendix:human_anno}
Following automated filtering, each candidate sample undergoes a rigorous, multi-stage human annotation and quality control process to ensure high data quality and annotation consistency. This process is as follows:

\begin{enumerate}[label=(\arabic*), leftmargin=*]
    \item \textbf{Systematic Training:} All annotators received detailed written guidelines and completed a trial annotation of 10 samples. These trials were meticulously reviewed by experts to ensure a unified understanding of the criteria.

    \item \textbf{Inter-annotator Cross-validation:} 
    \begin{enumerate}[label=(\roman*), leftmargin=*]
        \item \textit{Initial Annotation:} A sample is first annotated by Annotator A. The annotation content includes:
        \begin{itemize}
            \item {For Temporal Reasoning:} Task compliance checks, segment boundary delineation, textual descriptions for sub-clips and the global audio, scene classification, and audio quality scoring.
            \item {For Spatial Reasoning:}  Selecting appropriate segments, task classification, and the generation of a question, the correct answer, and distractor options for the multiple-choice format.
        \end{itemize}
        
        \item \textit{Review and Flagging:} The annotated sample is then fully reviewed by Annotator B, who flags any inconsistencies with detailed comments and marks the sample as ``failed''.
        
        \item \textit{Consensus through Negotiation:} Annotators A and B then discuss all flagged issues to reach a consensus and apply corrections. During the discussions, primary sources of ambiguity are as follows:
        \begin{itemize}
            \item {For Temporal Reasoning:} (a) The reasonableness of the segmented clip boundaries. (b) The existence of multiple logically plausible orderings for the segmented clips. (c) Significant discrepancies in audio quality scores.(d) Adherence to formatting and content guidelines for the captions.
            \item {For Spatial Reasoning:} (a) Whether the spatial perception presented in the audio unambiguously aligns with the annotated answer. (b) Whether the constructed question-answer pair clearly necessitates the use of audio spatial cues for resolution. (c) Potential ambiguity in mapping the event name mentioned in the question to a specific sound in the audio. (d) The appropriate difficulty and plausibility of the distractor options.
        \end{itemize}
        
        \item \textit{Expert Arbitration:} In the cases where a consensus cannot be reached, the sample is escalated to an expert panel for a final decision. If the experts still cannot agree, the sample is discarded.
    \end{enumerate}

    \item \textbf{Expert Spot-check:} After passing cross-validation, a random 10\% of samples undergo a final quality check by experts to ensure consistency and accuracy. Any discovered issues are then sent back for revision.
\end{enumerate}



\end{hlblock}

\section{Robust Evaluation}
\label[appendix]{appendix:robust_eval}
All questions in \ourbench are presented as clear multiple-choice questions with well-formatted options. We adopt classification accuracy as the evaluation metric. To determine the correctness of a response, we employ string matching to extract either the chosen option label (e.g., $<$A$>$) or the full text content of the option from the model's output. 

Furthermore, we implement a robust evaluation strategy to ensure rigorous and reliable results. For perception and spatial tasks, we adopt the CircularEval method from MM-Bench \citep{liu2024mmbench}. Specifically, each question is presented to the model $N$ times ($N$ is the number of options), with the option order cyclically rotated in each run to mitigate potential positional biases. For temporal tasks, we conduct three runs per question with different temporal segment orders to evaluate the model’s robustness to sequence variations.
Note that due to the significant API costs, GPT-4o Audio was evaluated only once per question.
This strategy yields two key metrics:  Average Accuracy (AA), the mean accuracy across all evaluation runs, and All-Correct Rate (ACR), the proportion of questions answered correctly in every single run, which serves as a stronger indicator of model reliability.

For models that do not support multi-audio input (only Audio Flamingo 3 and its Think variant among the models we evaluated), we concatenate the audios with a 2-second silence and specify this in the prompt. In contrast, for models that support multiple audio inputs, we feed them sequentially with textual indices.

To establish a human performance baseline, we conduct a human evaluation on a randomly sampled subset of approximately 10\% of the data from each task. This evaluation is performed by 10 university students, from whom we explicitly exclude anyone involved in data annotation or with domain-specific expertise, thereby ensuring a general, non-expert perspective.


\section{Breakdown Results}
\label[appendix]{appendix:detail_res}
In this section, we present detailed results for perception, temporal reasoning, and spatial reasoning on \ourbench, as shown in \cref{tab:results_perception}, \cref{tab:results_temporal}, and \cref{tab:results_spatial}.

\section{Further Analysis and Discussion}
\subsection{High Output Instability and Concentrated Predictions}
\label[appendix]{appendix:instability}
The reliability of model outputs on our benchmark is notably low, as evidenced by the stark contrast between their Average Accuracy (AA) and All-Correct-Rate (ACR) scores. Even the top-performing model, Gemini 2.5 Pro, exhibits an average drop of 25.01 percentage points from its AA to its ACR. This issue is even more pronounced for the majority of open-source models, which record an ACR near zero. This score indicates a complete failure to maintain consistent predictions under minor input perturbations. For these models, the instability often manifests as a tendency to concentrate predictions on a specific option, suggesting a reliance on superficial biases rather than genuine understanding.

\subsection{Ablation Study on Spatial Reasoning.}
\label[appendix]{appendix:spatial_ablation}
As shown in \cref{tab:results_spatial}, the results reveal a fundamental limitation of LALMs' in spatial perception. The \textbf{native input} inherently discards part of the multi-channel information during model preprocessing, which leads to a significant loss of spatial cues that are essential for fine-grained reasoning. On the other hand, the \textbf{channel-wise input} explicitly presents each channel with textual instructions, mitigating some of the information loss. 
Despite this, most existing models are not trained to handle multi-audio inputs. As a result, they consistently struggle to align channel representations and fail to make reliable use of interaural differences.
Overall, the pronounced gap between human and model performance highlights that spatial reasoning in audio remains an unsolved challenge, underscoring the need for audio encoders that natively support multi-channel audio input.

\begin{table}[htbp]
\centering
\caption{Results for the foundational perception task. Each cell reports \aaacr{AA}{ACR}: Average Accuracy (AA; overall accuracy across all runs) / All-Correct Rate (ACR; proportion of samples that are correct on every run). The best model in each category is shown in \textbf{bold}, and the second best is \underline{underlined}. }
\label{tab:results_perception}

\resizebox{\linewidth}{!}{%
\begin{tabular}{l c *{11}{c}}
\toprule
\multirow{2}{*}{\textbf{Model}} & \multirow{2}{*}{\textbf{Size}} &
\multicolumn{4}{c}{\textbf{Absolute Perception Range}} &
\multicolumn{6}{c}{\textbf{Relative Discrimination Sensitivity}} &
\multirow{2}{*}{\textbf{MA (\%)}} \\
\cmidrule(lr){3-6}\cmidrule(lr){7-12}
& & \textbf{Pitch\&Loudness} & \textbf{Azimuth} & \textbf{Elevation} & \textbf{Distance} &
\textbf{Pitch} & \textbf{Loudness} & \textbf{Duration} &\textbf{Azimuth} & \textbf{Elevation} & \textbf{Distance}  & \\
\midrule
Random Guess & \na &
  \aaacr{25.00}{0.39} & \aaacr{20.00}{0.03} & \aaacr{25.00}{0.39} & \aaacr{25.00}{0.39} &
  \aaacr{25.00}{0.39} & \aaacr{25.00}{0.39} & \aaacr{25.00}{0.39} & \aaacr{33.33}{3.7} &
  \aaacr{25.00}{0.39} & \aaacr{25.00}{0.39} & \aaacr{25.33}{0.68} \\ 
Human & \na &
  \aaacr{98.67}{\na} & \aaacr{73.33}{\na} & \aaacr{66.67}{\na} & \aaacr{70.00}{\na} &
  \aaacr{83.33}{\na} & \aaacr{85.56}{\na} & \aaacr{83.33}{\na} & \aaacr{83.33}{\na} &
  \aaacr{38.09}{\na} & \aaacr{73.68}{\na} &  \aaacr{75.60}{\na}\\ 
\midrule
SALMONN & 13B &  \aaacr{14.34}{0.00}
   & \aaacr{25.83}{0.63} & \aaacr{35.76}{0.00} & \aaacr{33.33}{0.00} &
  \aaacr{31.04}{0.00} & \aaacr{25.00}{0.00} & \aaacr{28.54}{0.00} & \aaacr{31.39}{3.89} &
  \aaacr{24.15}{0.00} & \aaacr{12.77}{0.00} & \aaacr{26.22}{0.45} \\ 
Audio Flamingo 3 & 8.4B &
  \aaacr{37.59}{0.00} & \aaacr{\textbf{27.92}}{3.13} & \aaacr{28.82}{0.00} & \aaacr{32.84}{0.00} &
   \aaacr{42.50}{1.67} & \aaacr{28.96}{0.00} & \aaacr{34.79}{0.00} &  \aaacr{38.61}{6.67} & \aaacr{\textbf{33.90}}{0.00} & \aaacr{35.56}{0.00} & \aaacr{34.15}{1.15} \\ 
Audio Flamingo 3 think & 8.4B  &\aaacr{51.75}{6.99}   & \aaacr{8.75}{0.00} & \aaacr{33.33}{1.04} & \aaacr{8.33}{0.00} &\aaacr{36.04}{8.33} &\aaacr{\textbf{45.63}}{2.50} & \aaacr{59.38}{38.33} & \aaacr{\underline{41.11}}{4.17} & \aaacr{12.29}{0.00} & \aaacr{10.00}{0.00} & \aaacr{30.66}{6.14}\\
Qwen2-Audio-Instruct & 8.4B &
  \aaacr{35.66}{1.40} & \aaacr{22.50}{0.00} & \aaacr{\textbf{48.61}}{10.76} & \aaacr{12.75}{0.98} &
  \aaacr{35.63}{0.00} & \aaacr{16.25}{0.00} & \aaacr{26.46}{0.00} & \aaacr{35.00}{8.06} &
  \aaacr{21.61}{1.69} & \aaacr{23.88}{0.00} & \aaacr{27.84}{2.29} \\
DeSTA2.5-Audio & 8.8B &
  \aaacr{16.96}{0.00} & \aaacr{21.25}{0.42} & \aaacr{45.49}{1.39} & \aaacr{35.78}{1.47} &
  \aaacr{11.67}{0.00} & \aaacr{11.25}{0.00} & \aaacr{22.71}{0.00} & \aaacr{33.06}{7.78} &
  \aaacr{10.59}{0.00} & \aaacr{29.44}{0.00} & \aaacr{23.82}{1.11} \\
BAT & 7B &
  \aaacr{0.00}{0.00} & \aaacr{\underline{26.04}}{26.04} & \aaacr{41.67}{41.67} & \aaacr{23.53}{23.53} &
  \aaacr{0.00}{0.00} & \aaacr{0.00}{0.00} & \aaacr{0.00}{0.00} & \aaacr{37.50}{37.50} &
  \aaacr{0.00}{0.00} & \aaacr{0.00}{0.00} & \aaacr{12.87}{12.87} \\
Phi4-MM & 5.5B &
  \aaacr{9.44}{0.00} & \aaacr{24.17}{0.00} & \aaacr{15.97}{0.00} & \aaacr{26.96}{0.00} &
  \aaacr{24.38}{0.00} & \aaacr{30.00}{0.00} & \aaacr{27.92}{0.00} & \aaacr{36.94}{0.00} &
  \aaacr{32.62}{0.00} & \aaacr{27.22}{0.00} & \aaacr{25.56}{0.00} \\
Kimi-Audio & 7B &
  \aaacr{18.71}{0.00} & \aaacr{18.12}{0.00} & \aaacr{38.19}{0.00} & \aaacr{18.13}{0.00} &
  \aaacr{24.38}{0.00} & \aaacr{32.29}{0.00} & \aaacr{34.17}{0.83} & \aaacr{39.72}{3.89} &
  \aaacr{25.00}{0.85} & \aaacr{9.44}{0.00} & \aaacr{25.82}{0.56} \\
MiDashengLM & 7B &
  \aaacr{48.95}{33.57} & \aaacr{20.63}{0.00} & \aaacr{\underline{48.26}}{11.81} & \aaacr{29.90}{0.98} &
  \aaacr{40.00}{34.17} & \aaacr{17.08}{0.83} & \aaacr{23.54}{7.50} & \aaacr{34.72}{8.61} &
  \aaacr{27.12}{1.69} & \aaacr{42.22}{6.11} & \aaacr{33.24}{10.53} \\
Step-Audio-2-mini & 7B &
  \aaacr{37.59}{0.00} & \aaacr{20.00}{0.00} & \aaacr{31.60}{0.69} & \aaacr{29.41}{0.00} &
  \aaacr{25.00}{0.00} & \aaacr{29.17}{0.00} & \aaacr{32.29}{0.00} & \aaacr{20.00}{0.00} &
  \aaacr{31.36}{0.00} & \aaacr{25.00}{0.00} & \aaacr{28.14}{0.07} \\
Gemma-3n-E4B-it & 7.5B &
  \aaacr{7.18}{0.00} & \aaacr{24.38}{4.17} & \aaacr{25.00}{0.00} & \aaacr{17.65}{0.00} &
  \aaacr{38.75}{0.00} & \aaacr{8.75}{0.00} & \aaacr{15.00}{5.83} & \aaacr{40.56}{1.94} &
  \aaacr{23.73}{0.00} & \aaacr{23.33}{0.00} & \aaacr{22.43}{1.19} \\
Ming-Lite-Omni-1.5 & 18.9B & \aaacr{28.67}{0.00} & \aaacr{20.21}{0.00} & \aaacr{27.78}{0.35} & \aaacr{30.39}{3.92} & \aaacr{16.67}{16.67} & \aaacr{16.67}{16.67} & \aaacr{16.67}{16.67} & \aaacr{\textbf{41.67}}{0.28} &
  \aaacr{\underline{32.81}}{0.00} & \aaacr{36.11}{0.00} &\aaacr{26.77}{5.46}  \\  
Qwen-2.5-Omni & 7B &
  \aaacr{27.45}{3.50} & \aaacr{18.33}{0.21} & \aaacr{27.57}{1.47} & \aaacr{\textbf{41.67}}{1.47} &
  \aaacr{48.13}{35.00} & \aaacr{39.79}{15.00} & \aaacr{38.33}{26.67} & \aaacr{16.11}{0.28} &
  \aaacr{11.02}{0.00} & \aaacr{40.56}{2.78} & \aaacr{30.90}{8.64} \\
Xiaomi-MiMo-Audio & 7B & \aaacr{36.71}{5.59} & \aaacr{18.54}{19.17} & \aaacr{48.26}{3.82} & \aaacr{36.27}{2.94} & \aaacr{46.04}{24.17} &\aaacr{36.46}{0.83} & \aaacr{17.70}{16.67} & \aaacr{40.56}{2.22} & \aaacr{20.98}{0.00} & \aaacr{27.78}{1.67} & \aaacr{32.93}{7.71}
 \\
Xiaomi-MiMo-Audio-think & 7B & \aaacr{43.01}{14.69} & \aaacr{11.67}{0.00} & \aaacr{25.69}{0.00} & \aaacr{39.21}{4.90} & \aaacr{28.13}{3.33} &\aaacr{15.21}{1.67} & \aaacr{22.71}{1.67} & \aaacr{29.44}{2.50} & \aaacr{21.88}{0.45} & \aaacr{32.22}{1.67} & \aaacr{26.92}{3.09}
   \\
MiniCPM-O-v2.6 & 8B & \aaacr{46.33}{8.39} & \aaacr{24.58}{0.21} & \aaacr{23.26}{0.35} & \aaacr{29.90}{0.00} & \aaacr{38.13}{3.33} &\aaacr{38.96}{4.17} & \aaacr{32.08}{3.33} & \aaacr{37.22}{2.78} & \aaacr{22.10}{0.22} & \aaacr{22.78}{0.00} & \aaacr{31.53}{2.28}
   \\
\midrule
GPT-4o Audio & \na &
  \aaacr{45.28}{\na} & \aaacr{16.67}{\na} & \aaacr{44.44}{\na} & \aaacr{3.92}{\na} &
  \aaacr{43.33}{\na} & \aaacr{36.04}{\na} & \aaacr{46.46}{\na} & \aaacr{29.58}{\na} &
  \aaacr{11.86}{\na} & \aaacr{40.00}{\na} & \aaacr{31.76}{\na} \\
Gemini 2.5 Flash & \na &
  \aaacr{\underline{62.59}}{18.19} & \aaacr{12.50}{0.00} & \aaacr{18.06}{0.35} & \aaacr{40.69}{1.47} &
  \aaacr{\underline{48.54}}{21.67} & \aaacr{\underline{40.83}}{6.67} & \aaacr{\underline{63.13}}{27.50} & \aaacr{37.08}{9.17} &
  \aaacr{25.42}{0.85} & \aaacr{\underline{48.33}}{4.44} & \aaacr{\underline{39.72}}{9.03} \\
Gemini 2.5 Pro & \na &
  \aaacr{\textbf{86.71}}{62.94} & \aaacr{25.83}{1.25} & \aaacr{5.88}{0.00} & \aaacr{\underline{41.18}}{5.88} &
  \aaacr{\textbf{63.33}}{52.50} & \aaacr{33.75}{15.83} & \aaacr{\textbf{78.96}}{68.33} & \aaacr{37.08}{13.75} &
  \aaacr{29.24}{6.36} & \aaacr{\textbf{64.44}}{12.22} & \aaacr{\textbf{46.64}}{23.91} \\
\bottomrule
\end{tabular}}
\end{table}

\begin{table}[htbp]
\centering
\caption{Results for the temporal reasoning task. Each cell reports \aaacr{AA}{ACR}: Average Accuracy (AA; overall accuracy across all runs) / All-Correct Rate (ACR; proportion of samples that are correct on every run). The best model in each category is shown in \textbf{bold}, and the second best is \underline{underlined}. }
\label{tab:results_temporal}

\resizebox{\linewidth}{!}{%
\begin{tabular}{l c *{6}{c}}
\toprule
\multirow{2}{*}{\textbf{Model}} & \multirow{2}{*}{\textbf{Size}} &
\multicolumn{2}{c}{\textbf{Continuous Processes}} &
\multicolumn{3}{c}{\textbf{Discrete Event Sequences}} &
\multirow{2}{*}{\textbf{OA (\%)}} \\
\cmidrule(lr){3-4}\cmidrule(lr){5-7}
& & \textbf{Object Spatial Motion} & \textbf{In-Situ State Evolution} & \textbf{Tool \& Appliance Operation} & \textbf{Daily Scene Scripts} &{\textbf{Event-Triggered Consequences}}
 & \\
\midrule
Random Guess & \na &
  \aaacr{14.29}{0.00} & \aaacr{14.29}{0.00} & \aaacr{14.29}{0.00} & \aaacr{14.29}{0.00} &
  \aaacr{14.29}{0.00}  & \aaacr{14.29}{0.00}\\
Human & \na &
  \aaacr{91.11}{\na} & \aaacr{88.89}{\na} & \aaacr{87.88}{\na} & \aaacr{83.33}{\na} &
  \aaacr{83.33}{\na}  & \aaacr{88.00}{\na} \\
\midrule
SALMONN & 13B &
  \aaacr{13.88}{0.74} & \aaacr{16.12}{0.00} & \aaacr{13.56}{1.96} & \aaacr{13.15}{1.11} &
   \aaacr{12.50}{0.00} & \aaacr{14.15}{0.89} \\
Audio Flamingo 3 & 8.4B & 
   \aaacr{8.55}{0.00} & \aaacr{10.08}{0.47} & \aaacr{8.66}{0.98} & \aaacr{7.22}{1.11} &
   \aaacr{8.33}{3.13} & \aaacr{8.67}{0.67} \\
Audio Flamingo 3 think & 8.4B & 
    \aaacr{14.37}{0.00} & \aaacr{11.78}{0.93} & \aaacr{15.36}{1.47} & \aaacr{12.96}{2.22} &
   \aaacr{11.46}{0.00} & \aaacr{13.59}{1.00} \\
Qwen2-Audio-Instruct & 8.4B &
  \aaacr{12.89}{0.00} & \aaacr{13.80}{0.93} & \aaacr{12.09}{0.00} & \aaacr{12.22}{1.11} &
  \aaacr{11.46}{0.00} & \aaacr{12.74}{0.44} \\
DeSTA2.5-Audio & 8.8B &
  \aaacr{16.98}{0.37} & \aaacr{15.97}{1.40} & \aaacr{19.93}{1.47} & \aaacr{15.56}{0.56} &
   \aaacr{11.46}{0.00} & \aaacr{16.93}{0.89} \\
BAT & 7B &
  \aaacr{0.00}{0.00} & \aaacr{0.00}{0.00} & \aaacr{0.00}{0.00} & \aaacr{0.00}{0.00} &
   \aaacr{0.00}{0.00} & \aaacr{0.00}{0.00} \\
Phi4-MM & 5.5B &
  \aaacr{17.72}{0.00} & \aaacr{15.50}{0.47} & \aaacr{16.34}{0.98} & \aaacr{17.04}{3.89} &
   \aaacr{20.83}{3.13} & \aaacr{16.85}{1.22} \\
Kimi-Audio & 7B &
  \aaacr{18.71}{1.49} & \aaacr{21.55}{2.33} & \aaacr{18.63}{0.49} & \aaacr{15.19}{2.22} &
   \aaacr{14.58}{0.00} & \aaacr{18.52}{1.56} \\
MiDashengLM & 7B &
  \aaacr{17.10}{0.37} & \aaacr{13.33}{0.00} & \aaacr{17.16}{1.96} & \aaacr{16.67}{2.22} &
   \aaacr{21.88}{0.00} & \aaacr{16.30}{1.00} \\
Step-Audio-2-mini & 7B &
  \aaacr{16.11}{0.37} & \aaacr{14.42}{0.00} & \aaacr{15.52}{0.00} & \aaacr{16.30}{0.00} &
   \aaacr{15.63}{0.00} & \aaacr{15.59}{0.11} \\
Gemma-3n-E4B-it & 7.5B &
  \aaacr{17.10}{0.00} & \aaacr{16.59}{0.00} & \aaacr{17.81}{0.00} & \aaacr{13.70}{0.00} &
   \aaacr{20.83}{0.00} & \aaacr{16.59}{0.00} \\
Ming-Lite-Omni-1.5 & 18.9B &
  \aaacr{17.47}{1.12} & \aaacr{16.59}{0.47} & \aaacr{13.89}{0.00} & \aaacr{17.59}{1.11} &
   \aaacr{14.58}{0.00} & \aaacr{16.37}{0.67} \\
Qwen-2.5-Omni & 7B &
  \aaacr{17.10}{0.37} & \aaacr{15.35}{0.93} & \aaacr{19.77}{1.47} & \aaacr{16.48}{0.56} &
   \aaacr{11.46}{0.00} & \aaacr{16.96}{0.78} \\
Xiaomi-MiMo-Audio & 7B & 
  \aaacr{18.22}{0.00} & \aaacr{18.14}{0.47} & \aaacr{17.16}{0.98} & \aaacr{20.19}{2.22} &
   \aaacr{26.04}{3.13} & \aaacr{18.63}{0.89} \\
Xiaomi-MiMo-Audio-think & 7B & 
  \aaacr{16.36}{0.37} & \aaacr{17.36}{0.47} & \aaacr{19.93}{1.96} & \aaacr{18.70}{2.22} &
   \aaacr{19.79}{0.00} & \aaacr{18.00}{1.11} \\
MiniCPM-O-v2.6 & 8B & 
  \aaacr{16.23}{0.00} & \aaacr{14.26}{0.93} & \aaacr{17.48}{0.49} & \aaacr{17.78}{0.56} &
   \aaacr{14.58}{0.00} & \aaacr{16.30}{0.44} \\
\midrule
GPT-4o Audio & \na &
  \aaacr{15.61}{\na} & \aaacr{16.28}{\na} & \aaacr{24.02}{\na} & \aaacr{22.78}{\na} &
   \aaacr{25.00}{\na} & \aaacr{19.44}{\na} \\
Gemini 2.5 Flash & \na &
  \aaacr{\underline{30.86}}{3.35} & \aaacr{\underline{23.41}}{3.72} & \aaacr{\underline{38.07}}{12.75} & \aaacr{\underline{30.19}}{7.22} &
   \aaacr{\underline{34.38}}{9.38} & \aaacr{\underline{30.70}}{6.56} \\
Gemini 2.5 Pro & \na &
  \aaacr{\textbf{63.82}}{38.66} & \aaacr{\textbf{43.72}}{17.67} & \aaacr{\textbf{69.77}}{46.08} & \aaacr{\textbf{57.22}}{38.33} &
   \aaacr{\textbf{48.96}}{28.13} & \aaacr{\textbf{58.52}}{34.89} \\
\bottomrule
\end{tabular}}
\end{table}

\begin{table}[htbp]
\centering
\caption{Results for the spatial reasoning task using native and channel-wise audio input. Each cell reports \aaacr{AA}{ACR}: Average Accuracy (AA; overall accuracy across all runs) / All-Correct Rate (ACR; proportion of samples that are correct on every run). The best model in each category is shown in \textbf{bold}, and the second best is \underline{underlined}. }
\label{tab:results_spatial}

\resizebox{\linewidth}{!}{%
\begin{tabular}{l c *{7}{cc}}
\toprule
\multirow{2}{*}{\textbf{Model}} & \multirow{2}{*}{\textbf{Size}} &
\multicolumn{2}{c}{\textbf{Single-Source Static Localization}} &
\multicolumn{2}{c}{\textbf{Multi-Source Spatial Relation}} &
\multicolumn{2}{c}{\textbf{Dynamic Trajectory Tracking}} &
\multicolumn{2}{c}{\textbf{OA (\%)}} \\
\cmidrule(lr){3-4}\cmidrule(lr){5-6}\cmidrule(lr){7-8} \cmidrule(lr){9-10}
& & \textbf{Native Input} & \textbf{Channel-wise Input} & \textbf{Native Input} & \textbf{Channel-wise Input} & \textbf{Native Input} & \textbf{Channel-wise Input} & \textbf{Native Input} & \textbf{Channel-wise Input} \\
\midrule
Random Guess & \na &
  \aaacr{33.33}{3.70} & \na & \aaacr{33.33}{3.70} & \na &
  \aaacr{33.33}{3.70} & \na  & \aaacr{33.33}{3.70} & \na\\
Human & \na &
  \aaacr{70.00}{\na} & \na & \aaacr{80.00}{\na} & \na &
  \aaacr{77.00}{\na} & \na  & \aaacr{73.72}{\na} & \na\\
\midrule
SALMONN & 13B &
  \aaacr{26.15}{3.18} & \aaacr{26.62}{3.18} & \aaacr{28.61}{4.42} &
  \aaacr{29.50}{5.31} & \aaacr{39.94}{0.94} & \aaacr{38.36}{0.94} & \aaacr{29.62}{2.99} & \aaacr{29.75}{3.19}\\
Audio Flamingo 3 & 8.4B &
  \aaacr{37.22}{1.77} & \aaacr{\textbf{42.87}}{2.12} & \aaacr{38.35}{4.42} & \aaacr{46.31}{10.62} &
  \aaacr{44.03}{4.72} & \aaacr{\underline{46.23}}{0.94} & \aaacr{38.91}{2.99} & \aaacr{\underline{44.35}}{3.78} \\
Audio Flamingo 3 think & 8.4B &
  \aaacr{35.45}{7.42} & \aaacr{\textbf{42.87}}{13.78} & \aaacr{37.46}{23.01} & \aaacr{46.02}{23.01} &
  \aaacr{38.05}{18.87} & \aaacr{37.11}{19.81} & \aaacr{36.45}{13.35} & \aaacr{42.36}{17.13} \\
Qwen2-Audio-Instruct & 8.4B &
  \aaacr{21.32}{8.48} & \aaacr{6.36}{1.77} & \aaacr{24.78}{3.54} & \aaacr{12.09}{4.42} &
  \aaacr{15.09}{0.94} & \aaacr{11.64}{2.83} & \aaacr{20.78}{5.78} & \aaacr{8.76}{2.59} \\
DeSTA2.5-Audio & 8.8B &
  \aaacr{23.67}{2.83} & \aaacr{20.38}{4.59} & \aaacr{34.81}{9.73} &
  \aaacr{41.30}{19.47} & \aaacr{37.74}{10.38} & \aaacr{32.08}{21.70} & \aaacr{29.15}{5.98} & \aaacr{27.56}{11.55} \\
BAT & 7B &
  \aaacr{0.00}{0.00} & \aaacr{0.00}{0.00} & \aaacr{0.00}{0.00} & \aaacr{0.00}{0.00} &
  \aaacr{0.00}{0.00} & \aaacr{0.00}{0.00} & \aaacr{0.00}{0.00} & \aaacr{0.00}{0.00}\\
Phi4-MM & 5.5B &
  \aaacr{33.10}{0.35} & \aaacr{32.63}{0.35} & \aaacr{27.14}{0.88} & \aaacr{29.79}{0.88} &
  \aaacr{34.28}{0.94} & \aaacr{33.02}{0.00} & \aaacr{32.01}{0.59} & \aaacr{32.07}{0.40} \\
Kimi-Audio & 7B &
  \aaacr{27.56}{3.53} & \aaacr{16.49}{3.53} & \aaacr{38.94}{15.04} & \aaacr{22.42}{8.85} &
  \aaacr{44.03}{7.55} & \aaacr{40.25}{8.49} & \aaacr{33.60}{6.97} & \aaacr{22.84}{5.77}\\
MiDashengLM & 7B &
  \aaacr{\textbf{43.11}}{15.19} & \aaacr{37.22}{17.67} & \aaacr{\underline{45.43}}{23.89} & \aaacr{42.77}{16.81} &
  \aaacr{\textbf{46.23}}{30.19} & \aaacr{45.60}{21.70} & \aaacr{\textbf{44.29}}{20.32} & \aaacr{40.24}{18.33}\\
Step-Audio-2-mini & 7B &
  \aaacr{33.33}{0.00} & \aaacr{33.33}{0.00} & \aaacr{31.27}{0.00} & \aaacr{37.46}{0.00} &
  \aaacr{37.74}{6.38} & \aaacr{35.22}{2.83} & \aaacr{33.80}{1.34} & \aaacr{34.66}{0.60}\\
Gemma-3n-E4B-it & 7.5B &
  \aaacr{23.32}{1.41} & \aaacr{28.27}{6.01} & \aaacr{41.89}{15.04} & \aaacr{36.58}{7.96} &
  \aaacr{33.96}{5.66} & \aaacr{40.57}{8.49} & \aaacr{29.75}{5.37} & \aaacr{32.74}{6.97}\\
Ming-Lite-Omni-1.5 & 18.9B &
  \aaacr{20.14}{6.36} & \aaacr{34.63}{6.01} & \aaacr{35.10}{9.73} & \aaacr{33.04}{9.73} &
  \aaacr{38.36}{18.87} & \aaacr{39.94}{20.75} & \aaacr{27.35}{9.76} & \aaacr{35.39}{9.96}\\
Qwen-2.5-Omni & 7B &
  \aaacr{39.46}{7.07} & \aaacr{36.98}{15.19} & \aaacr{41.30}{18.58} & \aaacr{35.10}{15.93} &
  \aaacr{27.04}{17.92} & \aaacr{34.59}{8.49} & \aaacr{37.25}{11.95} & \aaacr{36.05}{13.94}\\
Xiaomi-MiMo-Audio & 7B &
  \aaacr{36.16}{0.71} & \aaacr{41.58}{5.65} & \aaacr{41.30}{5.31} & \aaacr{38.05}{4.42} &
  \aaacr{\underline{45.28}}{9.43} & \aaacr{44.34}{9.43} & \aaacr{39.24}{3.58} & \aaacr{41.37}{6.17} \\
Xiaomi-MiMo-Audio-think & 7B &
  \aaacr{34.28}{7.42} & \aaacr{25.44}{2.83} & \aaacr{44.54}{14.16} & \aaacr{37.76}{7.96} &
  \aaacr{36.79}{7.55} & \aaacr{27.99}{3.77} & \aaacr{37.12}{8.96} & \aaacr{28.75}{4.18} \\
MiniCPM-O-v2.6 & 8B &
  \aaacr{29.92}{3.18} & \aaacr{27.92}{2.83} & \aaacr{43.36}{11.50} & \aaacr{39.53}{12.39} &
  \aaacr{38.36}{26.42} & \aaacr{35.53}{17.92} & \aaacr{34.73}{9.96} & \aaacr{32.14}{8.17} \\
\midrule
GPT-4o Audio & \na &
  \aaacr{\underline{41.81}}{\na} & \aaacr{\underline{42.76}}{\na} & \aaacr{43.07}{\na} & \aaacr{\textbf{54.87}}{\na} &
  \aaacr{39.94}{\na} & \aaacr{42.45}{\na} & \aaacr{41.70}{\na} & \aaacr{\textbf{45.42}}{\na}\\
Gemini 2.5 Flash & \na &
  \aaacr{24.62}{4.95} & \aaacr{40.75}{7.42} & \aaacr{43.07}{15.93} & \aaacr{43.07}{17.70} &
  \aaacr{22.64}{2.83} & \aaacr{40.57}{11.32} &  \aaacr{28.35}{6.97} & \aaacr{41.23}{10.56} \\
Gemini 2.5 Pro & \na &
  \aaacr{40.87}{10.95} & \aaacr{34.98}{11.66} & \aaacr{\textbf{48.97}}{25.66} & \aaacr{\underline{49.26}}{20.35} &
  \aaacr{\underline{45.28}}{14.15} & \aaacr{\textbf{47.17}}{7.55} & \aaacr{\underline{43.62}}{14.94} & \aaacr{40.77}{12.75} \\
\bottomrule
\end{tabular}}
\end{table}

\section{Case Study}
\begin{figure*}[tb!]
	\centering
	\includegraphics[width=\columnwidth, page=1]{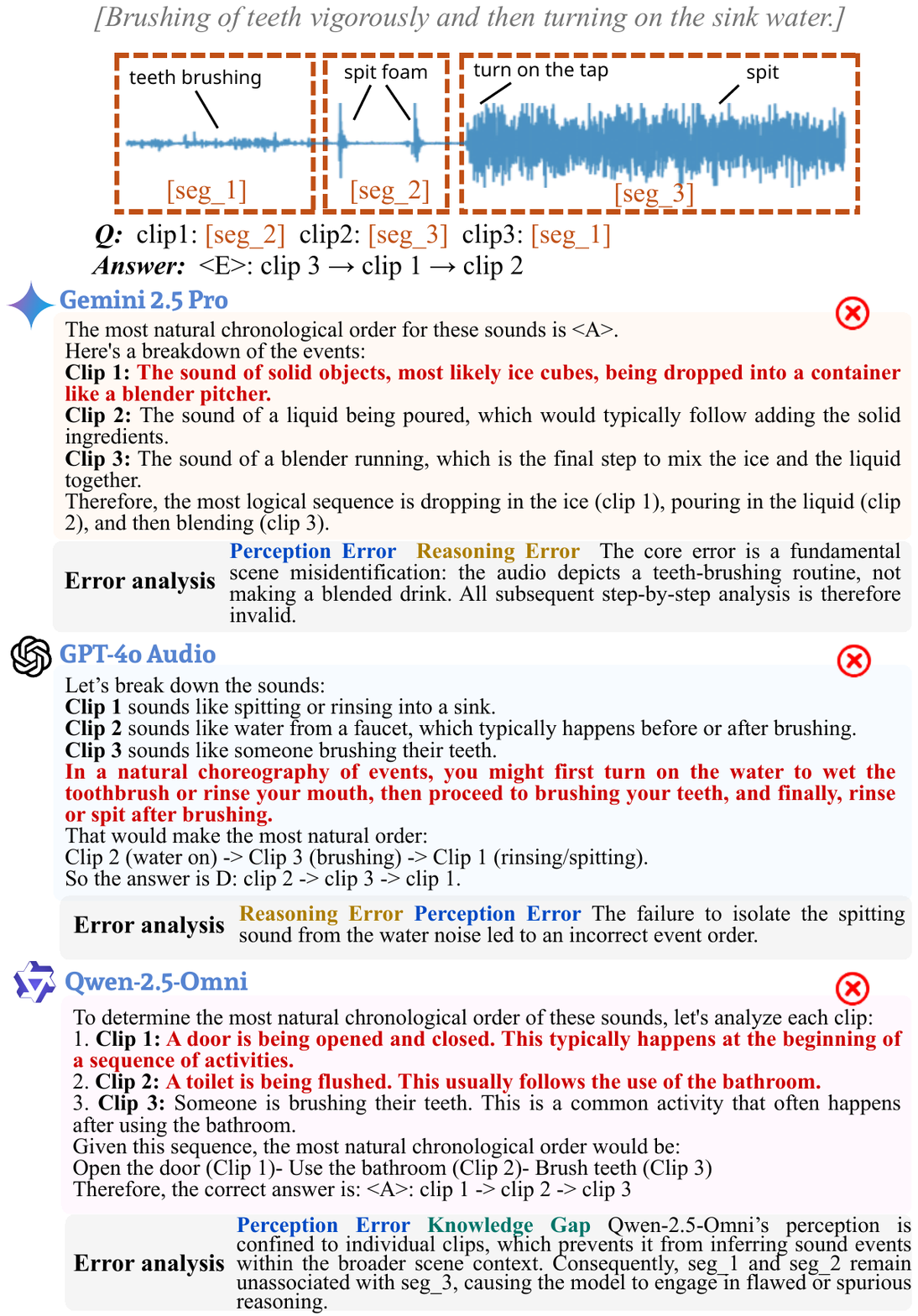}

	\caption{An error case from the temporal reasoning task.
    }
	\label{fig:temporal_case1}

\end{figure*}

\begin{figure*}[tb!]
	\centering
	\includegraphics[width=\columnwidth, page=2]{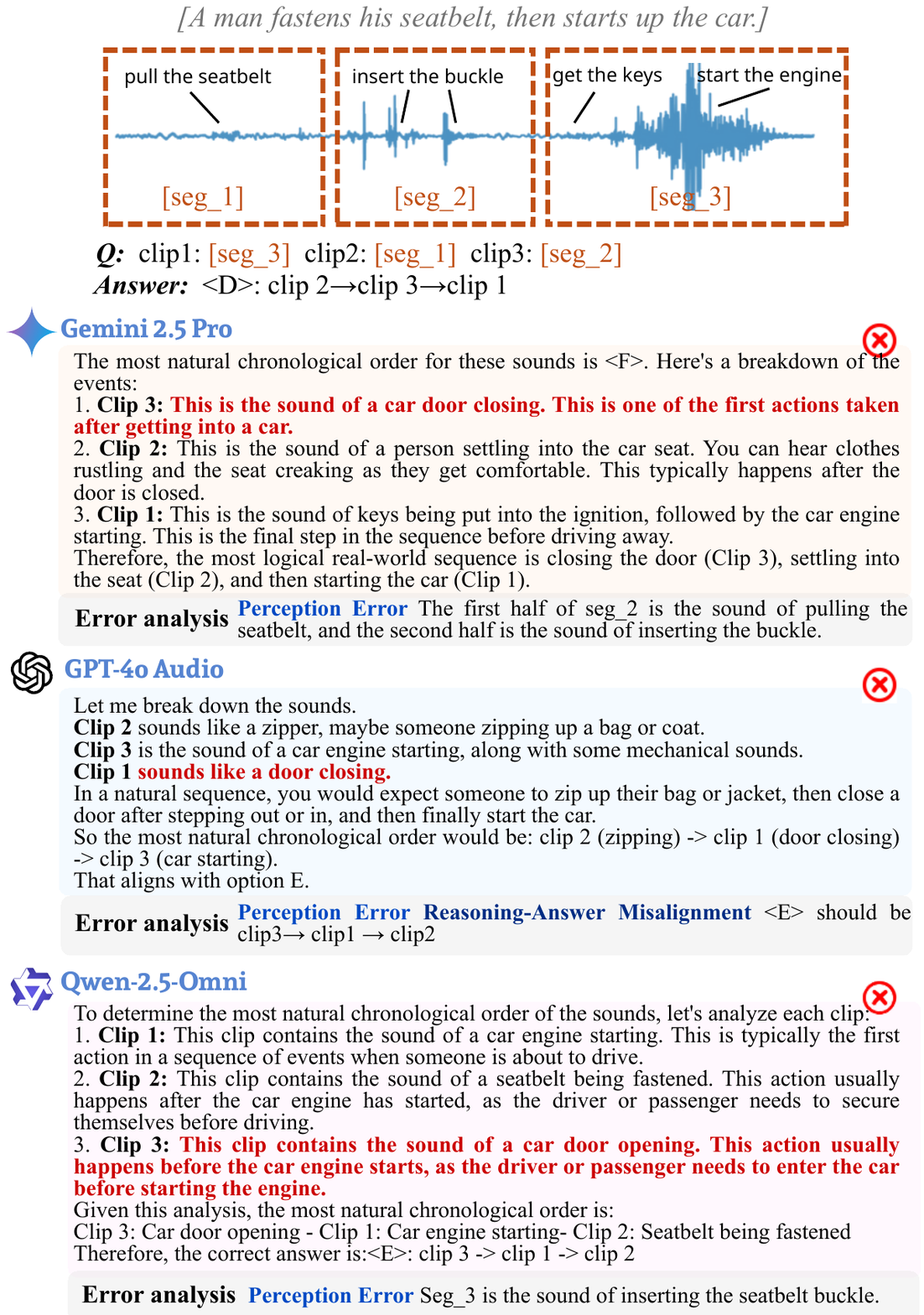}

	\caption{An error case from the temporal reasoning task.
    }
	\label{fig:temporal_case2}

\end{figure*}

\begin{figure*}[tb!]
	\centering
	\includegraphics[width=\columnwidth, page=3]{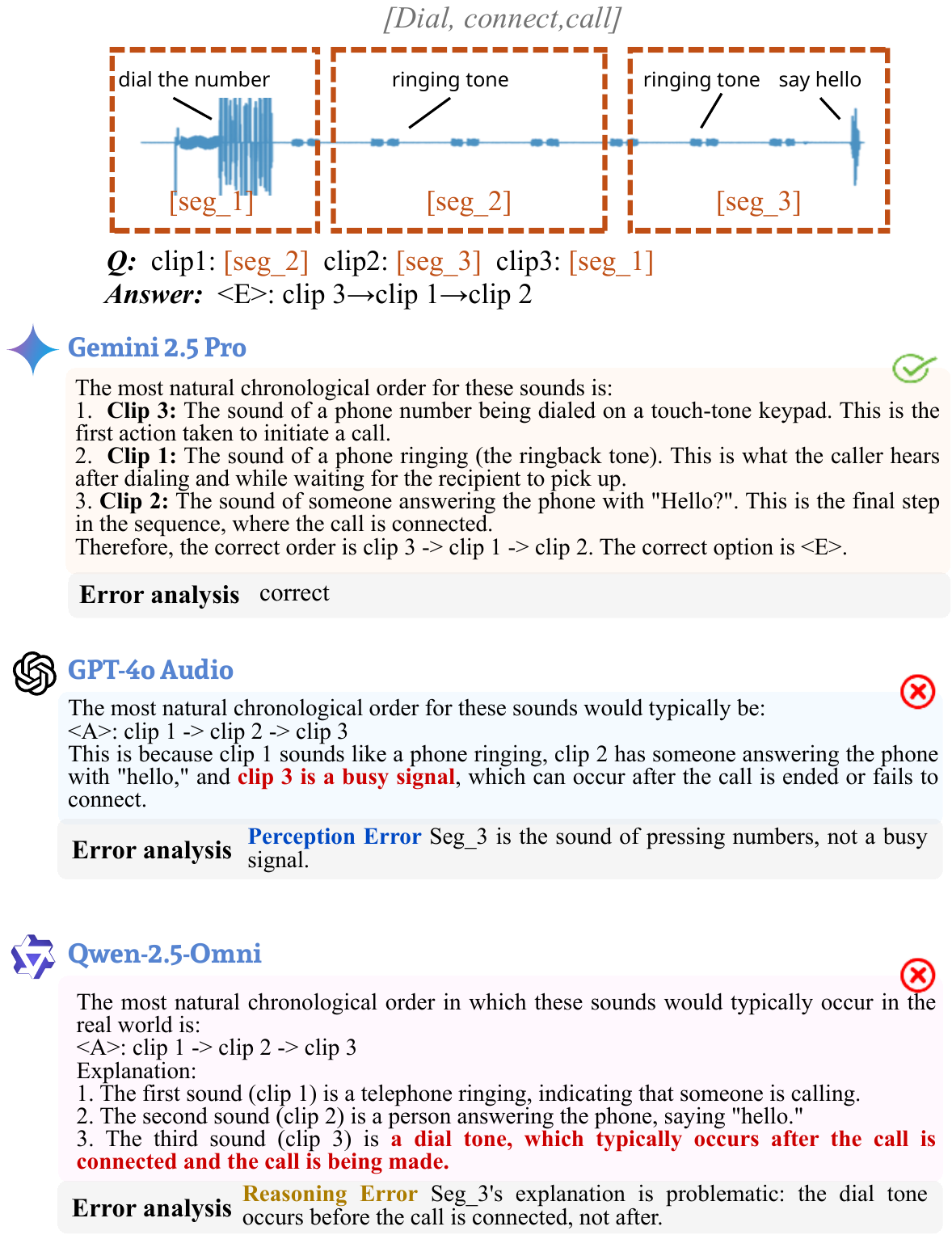}

	\caption{An error case from the temporal reasoning task.
    }
	\label{fig:temporal_case3}
   
\end{figure*}

\begin{figure*}[tb!]
	\centering
	\includegraphics[width=\columnwidth, page=4]{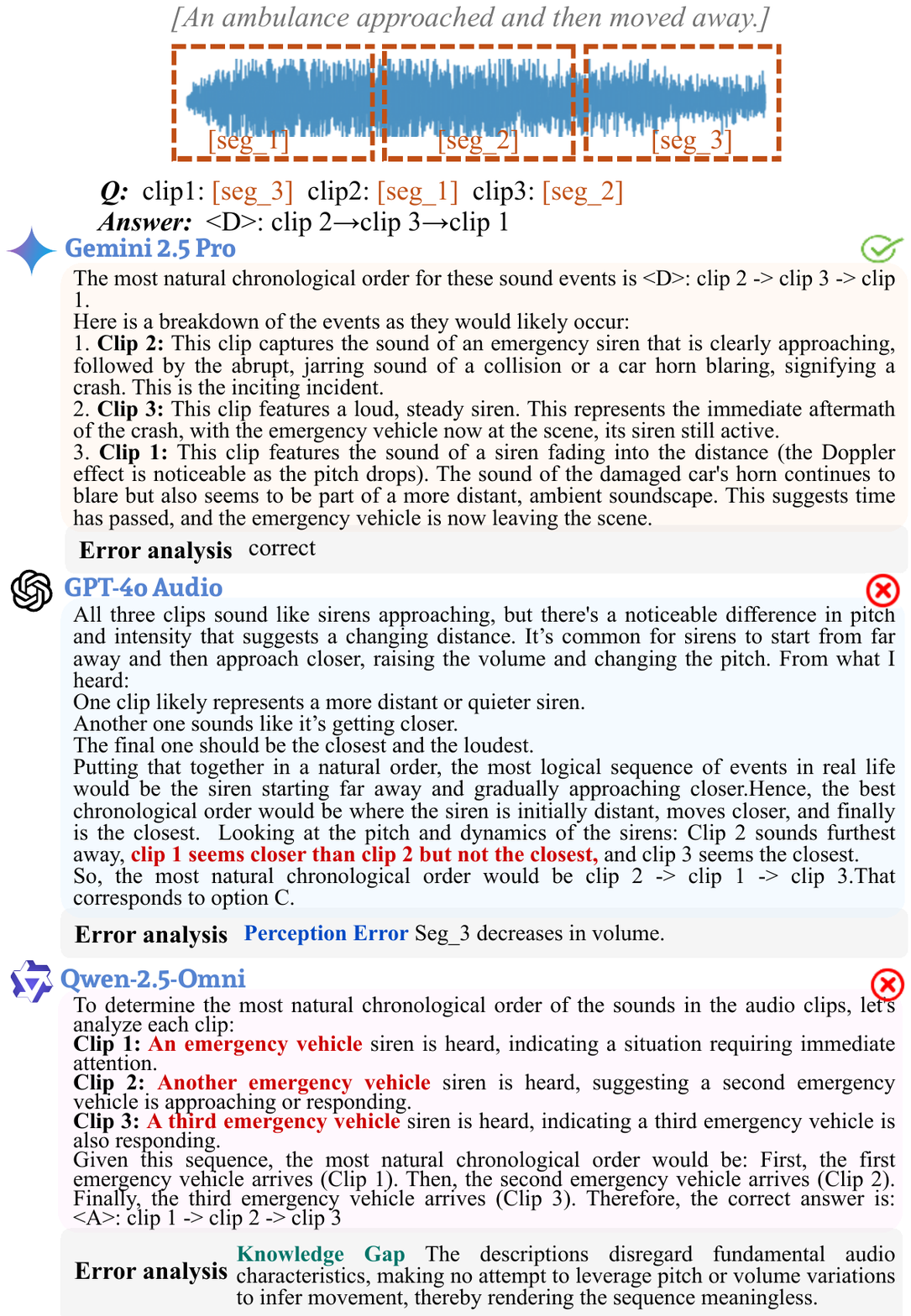}
   
	\caption{An error case from the temporal reasoning task.
    }
	\label{fig:temporal_case4}
   
\end{figure*}

\begin{figure*}[tb!]
	\centering
	\includegraphics[width=\columnwidth, page=5]{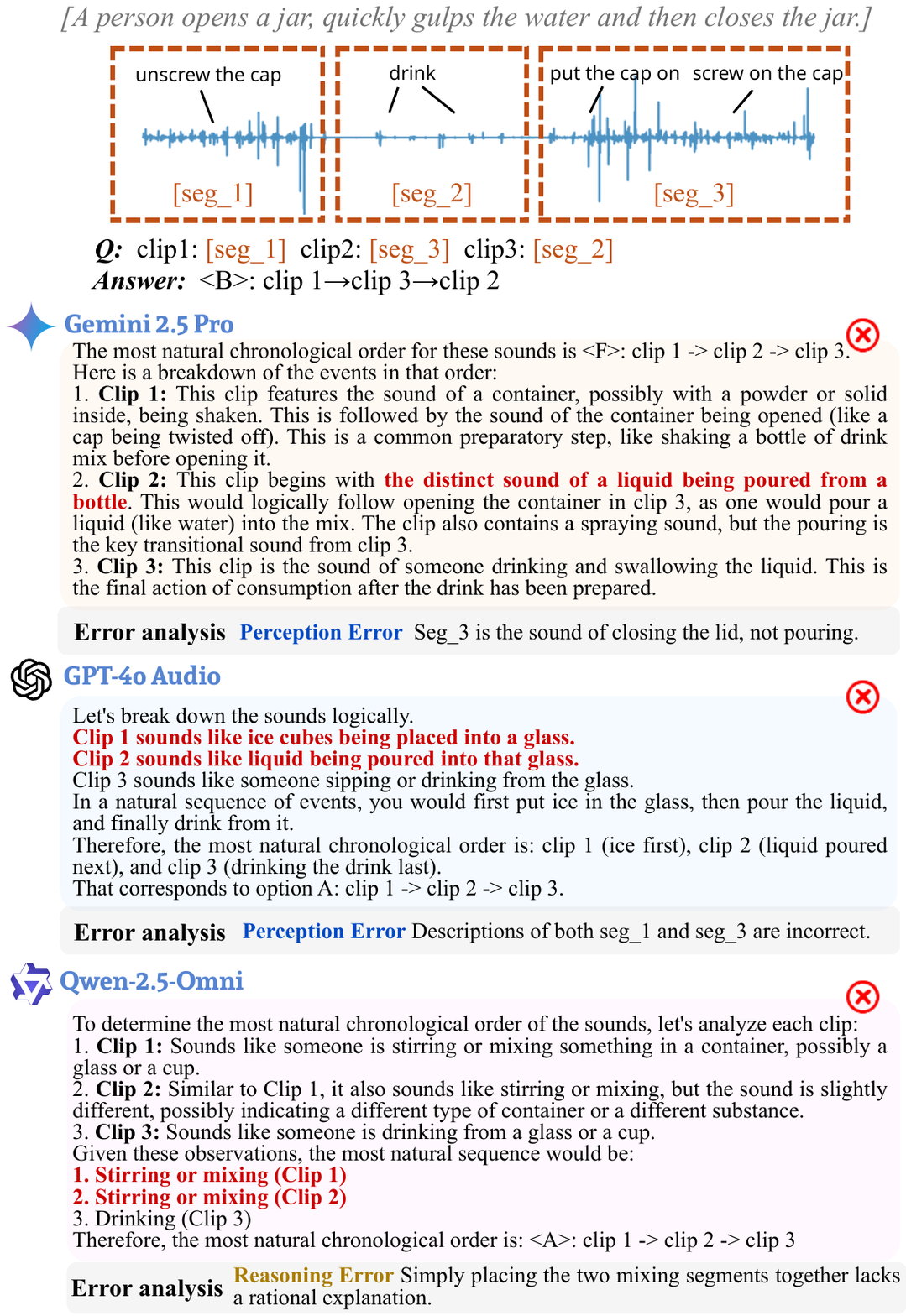}
    
	\caption{An error case from the temporal reasoning task.
    }
	\label{fig:temporal_case5}
    
\end{figure*}

\begin{figure*}[tb!]
	\centering
	\includegraphics[width=\columnwidth, page=6]{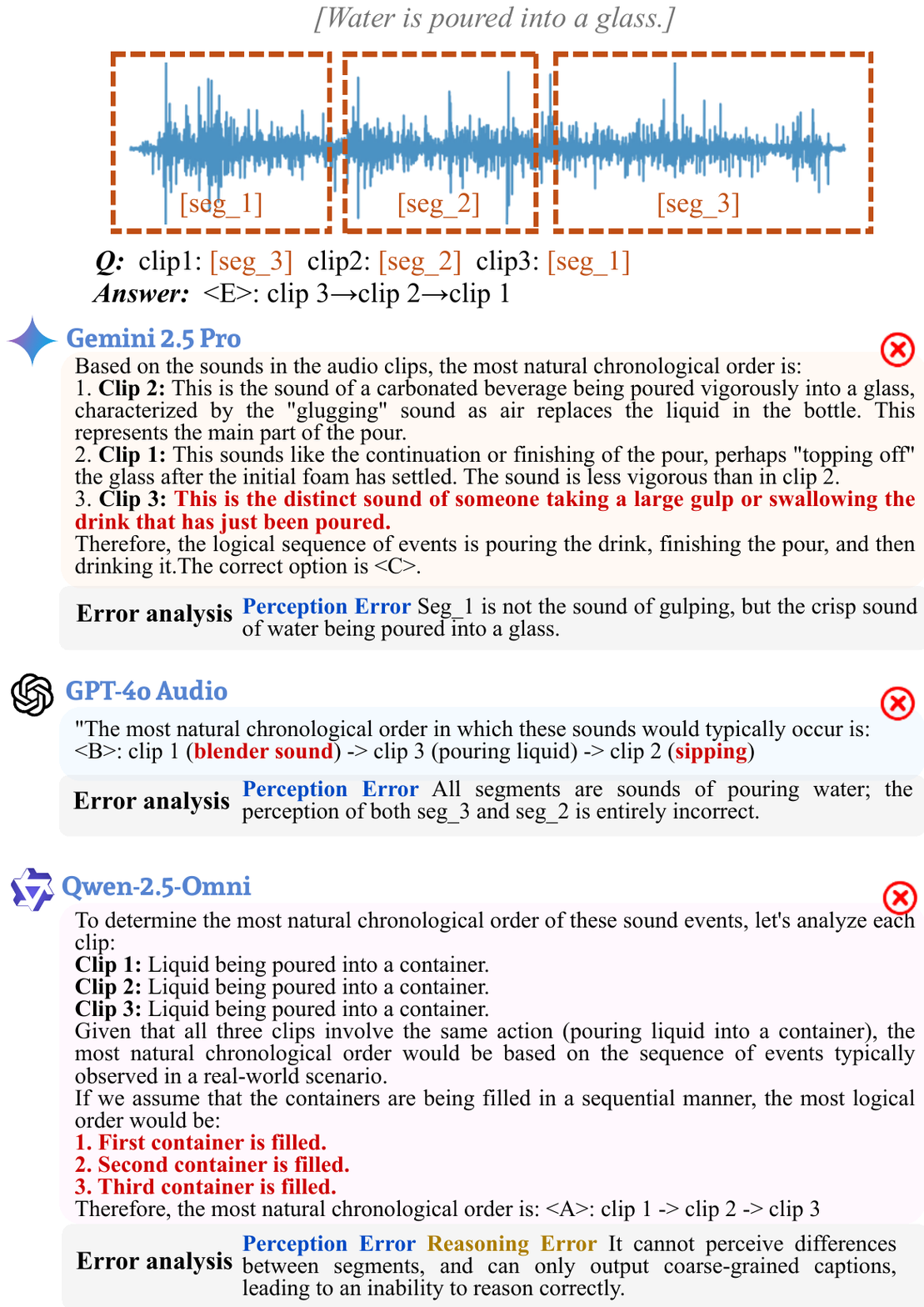}
    
	\caption{An error case from the temporal reasoning task.
    }
	\label{fig:temporal_case6}

\end{figure*}

\begin{figure*}[tb!]
	\centering
	\includegraphics[width=\columnwidth]{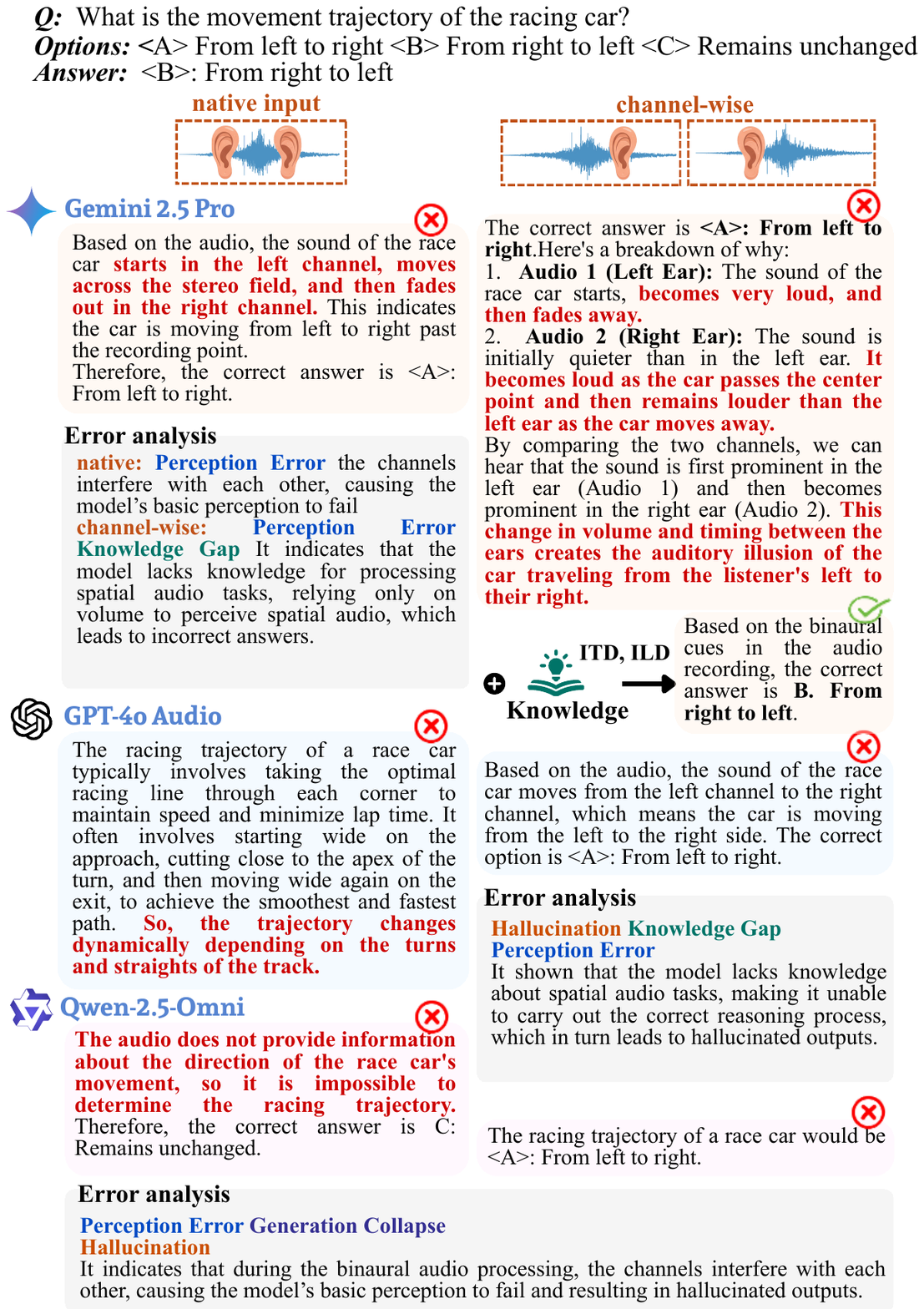}
    
	\caption{An error case from the spatial reasoning task.
    }
	\label{fig:spatial_case1}

\end{figure*}

\label[appendix]{appendix:cases}
In this section, we present several case studies of error analysis, including temporal reasoning (\cref{fig:temporal_case1,fig:temporal_case2,fig:temporal_case3,fig:temporal_case4,fig:temporal_case5,fig:temporal_case6}) and spatial reasoning (\cref{fig:spatial_case1}).

